\documentclass[12pt]{article}
\pdfoutput=1
\usepackage{jheppub}
\usepackage{subcaption}
\usepackage{xspace}
\usepackage{booktabs}
\usepackage{ulem}
\usepackage{listings}
\usepackage{braket}
\usepackage{url}
\usepackage{setspace}
\usepackage{color}

\usepackage[utf8]{inputenc}
\usepackage{amsmath,amssymb}
\usepackage{graphicx}
\newcommand{\be}{\begin{equation}}
\newcommand{\ee}{\end{equation}}
\newcommand{\bear}{\begin{eqnarray}}
\newcommand{\eear}{\end{eqnarray}}
\newcommand{\ba}{\begin{array}}
\newcommand{\ea}{\end{array}}

\newcommand{\Ap}{A^\prime}
\newcommand{\mAp}{m_{A^\prime}}

\title{Sensitivity of the SHiP experiment to light dark matter}
\collaborationImg{\includegraphics[width=.15\textwidth]{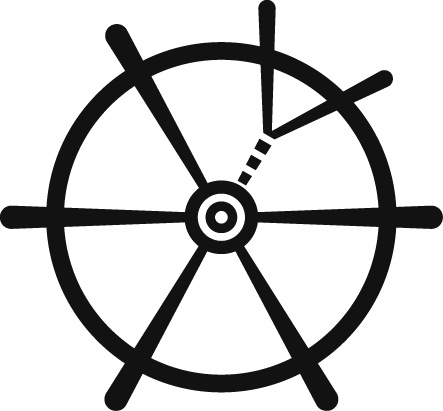}}
\collaboration{SHiP Collaboration}
\author{C.~Ahdida$^{44}$,
A.~Akmete$^{48}$,
R.~Albanese$^{14,d,h}$,
A.~Alexandrov$^{14,32,34,d}$,
A.~Anokhina$^{39}$,
S.~Aoki$^{18}$,
G.~Arduini$^{44}$,
E.~Atkin$^{38}$,
N.~Azorskiy$^{29}$,
J.J.~Back$^{54}$
% ...

\vspace*{1cm}

{\footnotesize \it

% ...
$ ^{14}$Sezione INFN di Napoli, Napoli, Italy\\
% ...
$ ^{18}$Kobe University, Kobe, Japan\\
% ...
$ ^{29}$Joint Institute for Nuclear Research (JINR), Dubna, Russia\\
% ...
$ ^{32}$P.N.~Lebedev Physical Institute (LPI RAS), Moscow, Russia\\
% ...
$ ^{34}$National University of Science and Technology ``MISiS``, Moscow, Russia\\
% ...
$ ^{38}$National Research Nuclear University (MEPhI), Moscow, Russia\\
$ ^{39}$Skobeltsyn Institute of Nuclear Physics of Moscow State University (SINP MSU), Moscow, Russia\\
% ...
$ ^{44}$European Organization for Nuclear Research (CERN), Geneva, Switzerland\\
% ...
$ ^{48}$Middle East Technical University (METU), Ankara, Turkey\\
% ...
$ ^{54}$University of Warwick, Warwick, United Kingdom\\
% ...
$ ^{d}$Universit\`{a} di Napoli ``Federico II``, Napoli, Italy\\
% ...
$ ^{h}$Consorzio CREATE, Napoli, Italy\\
% ...
}
}
\emailAdd{luca.buonocore@physik.uzh.ch}
\emailAdd{martina.ferrillo@cern.ch}

\begin{document}
\newpage
\abstract{
Dark matter is a well-established theoretical addition to the Standard Model supported by many observations in modern astrophysics and cosmology. In this context, the existence of weakly interacting massive particles represents an appealing solution to the observed thermal relic in the Universe. Indeed, a large experimental campaign is ongoing for the detection of such particles in the sub-GeV mass range. Adopting the benchmark scenario for light dark matter particles produced in the decay of a dark photon, with $\alpha_D=0.1$ and $~m_{A'}=3m_{\chi}$, we study the potential of the SHiP experiment to detect such elusive particles through its Scattering and Neutrino detector (SND). In its 5-years run, corresponding to $2\cdot 10^{20}$ protons on target from the CERN SPS, we find that SHiP will improve the current limits in the mass range for the dark matter from about 1~MeV to 300~MeV. In particular, we show that SHiP will probe the thermal target for Majorana candidates in most of this mass window and even reach the Pseudo-Dirac thermal relic.
}
\maketitle
\section{Introduction}

One of the main challenges in particle physics today is figuring out the microscopic identity and the cosmological origin of dark matter~(DM).
The theoretical landscape is broad and it spans over many orders of magnitude in the mass/coupling parameter space. A compelling idea to explore is DM as a thermal relic of the early universe. The canonical example of this scenario is the Weakly Interacting Massive Particle (WIMP), a particle in the GeV-TeV mass range interacting with the visible sector via weak-sized interactions. Searches for WIMPs are in full swing \cite{Schumann:2019eaa,Fox:2019bgz}: however, the interesting parameter space goes beyond that has been explored in the past decade: thermal DM can be as heavy as 100 TeV or as light as a few keV. Recently,  a lot of attention has been directed towards light DM~(LDM) in the keV-GeV mass range \cite{cosmicvision}.
\par
Direct detection has traditionally employed the Migdal Effect~\cite{Ibe:2017yqa} using liquid Argon~\cite{PhysRevLett.121.081307,Angloher:2011uu} or liquid Xenon~\cite{PhysRevLett.109.021301,Aprile:2018dbl,Macolino:2020uqq,Cui:2017nnn}, while a novel strategy based on silicon devices has allowed to design new experiments optimised for sub-GeV DM, as SENSEI~\cite{PhysRevLett.121.061803}.
Since current DM direct detection experiments searching for elastic nuclear recoils rapidly lose sensitivity once the candidate mass drops below a few GeV~\cite{Schumann:2019eaa,Roszkowski:2017nbc}, experiments at the intensity frontier represent an alternative yet appealing route and play an important role in this quest \cite{cosmicvision}.  
Fixed target experiments represent the prototype for such searches, although other collider experiments might be relevant in the same parameter space, as showed by the mono-photon searches at BaBar~\cite{Lees:2017lec} and Belle II~\cite{Kou:2018nap}.

In particular, neutrino fixed target experiments could efficiently search for LDM via signatures of DM scattering with electrons and/or nuclei in their near detectors~\cite{Batell:2009di, deNiverville:2011it, deNiverville:2012ij, Dharmapalan:2012xp, Batell:2014yra, Soper:2014ska,Dobrescu:2014ita,Coloma:2015pih,Frugiuele:2017zvx,nova,millicharged,MiniBooneE}.

Here we present the sensitivity of the SHiP scattering and neutrino detector (SND), to LDM. We focus on the hypothesis that the DM couples to the SM through the exchange of a massive vector mediator, dubbed in the literature dark photon, and we have considered the cleanest signature given by the LDM-electron scattering. The scattering with nuclei, both coherent and deep inelastic scattering, although plagued by a larger neutrino background, might be an alternative detection strategy and will be the subject of a forthcoming dedicated analysis. 

In a proton beam dump experiment signal yields are largely reduced as the interaction with the dark photon $\Ap$ is probed twice, if compared to electron fixed target experiments which make use of search strategies based on missing energy, such as NA64~\cite{NA64:2019imj}, or missing momentum, such as the LDMX proposal~\cite{Akesson:2018vlm}. Indeed, the LDM detection is achieved through its scattering within the downstream detector. Hence, the expected LDM yield scales as $\epsilon^4\,\alpha_D$ ($\epsilon$ being the interaction strength of the dark photon to SM particles and $\alpha_D$ the LDM-$\Ap$ coupling), where a factor $\epsilon^2$ comes from production and another $\epsilon^2 \alpha_D$ is due to detection. This has to be compared to the $\epsilon^2$ scaling of typical missing energy/momentum experiments, which prove however to be not sensitive to LDM coupling constant $\alpha_D$.
Due to their higher penetrating power and enhancements from meson decay reactions and/or strong interactions, proton beams are characterised by dark photon production rates larger than the ones achievable in electron beams of comparable intensity, which in part compensate for the detection suppression factor. 

The potential to directly probe the dark sector mediator coupling $\alpha_D$, together with a wider sensitivity which encompasses other viable dark matter models, shows to a large extent the complementarity between the two approaches. This is even a more pressing aspect in the light of a possible discovery,
as in general the observation of an excess alone is not sufficient to claim a discovery of a Dark matter particle. Indeed, intensity frontier probes do not depend on whether the particle $\chi$ produced through prompt DP decay is DM or not, as the only necessary ingredient is its stability concerning the target-detector distance. The observed excess might have an instrumental origin rather than a genuine New Physics (NP) effect.  
This applies also to the constraints that the SHiP experiment can place. With this regard, invaluable contribution could come from complementary DM observations from a cosmic source to unequivocally probe its thermal origin. 
In addition, since the SHiP experiment has a direct sensitivity to LDM interactions, we anticipate the possibility to use the time of flight measurement to uncontroversially discriminate massive NP particles from the SM neutrino background. 

The paper is organised as follows: in Section~\ref{sec:Vec_Portal} we give a brief presentation of the model focusing on the main motivations. After introducing the SHiP experiment in Section~\ref{sec:SHiP_Experiment}, we discuss the relevant production and detection mechanisms, in Section~\ref{sec:LDM_prod_detect}. The detailed analysis of the neutrino background is the topic of Section~\ref{sec:Bkg}. We then pass to the discussion of the signal reviewing the main processes taken into account in our simulation. Finally, we show the main results on the sensitivity reach of the SHiP experiment in Section~\ref{sec:Sensitivity} and we give our conclusions in Section~\ref{sec:Conclusions}.      
 \section{Vector Portal}
 \label{sec:Vec_Portal}
Thermal freeze-out can naturally explain the origin of the DM relic density for a sub-GeV particle provided the interaction with the visible sector is mediated by a new light force carrier~\cite{Fox:2019bgz,Batell:2017kty}.
Here, we will consider as benchmark model the dark photon (DP)~\cite{Holdom:1985ag}  vector portal where the DP  $\Ap_{\mu}$, is the gauge boson of a new dark gauge group $U(1)_D$ kinetically mixed with the photon, and a scalar $\chi$ charged under $U(1)_D$ that serves as a DM candidate. Then, the low-energy effective Lagrangian describing the DM reads
\be 
\mathcal{L}_{ \rm DM}=\mathcal{L}_{A^{\prime}}+\mathcal{L}_\chi 
\ee
where:
\be
\mathcal{L}_{A^{\prime}} =- \frac{1}{4} F'_{\mu \nu}F^{\prime \mu \nu} +\frac{m^2_{\Ap} }{2}A^{\prime \mu} A^{\prime}_{ \mu}-\frac{1}{2} \epsilon F^{\prime}_{\mu \nu} F^{\mu \nu},
\ee
where $\epsilon$ is the DP-photon kinetic mixing parameter and $m_{A^{\prime}}$ is the mass of the DP while:
 \be
\mathcal{L}_\chi =  \frac{i g_D}{2}  A^{\prime \mu} J_{\mu}^{\chi}+\frac{1}{2}  \partial_{\mu} \chi^\dagger \partial^{\mu} \chi - m_{\chi}^2 \chi^\dagger \chi,
\ee
where $ J_{\mu}^{\chi}=  \left[ (\partial_\mu \chi^\dagger) \chi  -  \chi^\dagger  \partial_\mu \chi \right]$, $g_D$ is the $U(1)_D$ gauge coupling and $m_{\chi}$ is the mass of the dark matter particle.
The region of the parameter space relevant for $\chi$ searches at beam-dump facilities corresponds to $ \mAp > 2 m_{\chi}$ and $ g_D \gg \epsilon e$ which implies  $ BR( A' \to \chi\chi^\dagger) \sim 1$. 
\par
In case $\chi$ is DM, precise measurements of the temperature anisotropies of the cosmic microwave background (CMB) radiation significantly constraint the parameter space. In particular, they rule out Dirac fermions with mass below 10 GeV as a thermal DM candidate and more in general every DM candidate that acquires its relic abundance via $s$-wave annihilation into SM particles~\cite{Lin:2011gj,Ade:2015xua}.
Hence, a complex scalar dark matter candidate $\chi$ is safe from such constraints as well as a Majorana or Pseudo-Dirac fermion. 
Tighter bounds come instead from the Planck measurement of the effective number of neutrino species  $N_{\mathrm{eff}}$ \cite{Ade:2015xua} and rule out the minimal DP model considered here if the complex scalar is lighter than 9 MeV~\cite{Depta:2019lbe}. 
\par
In order to show the region of parameter space relevant for thermal freeze-out, we will present the SHiP sensitivity in the $ (m_{\chi},Y)$  plane where $Y$ is defined as:\be
Y \equiv \epsilon^2 \alpha_D \left (\frac{ m_{\chi}}{\mAp}\right)^4, \qquad \alpha_D = \frac{g_D^2}{4\pi}.  
\ee
In the assumption $ \mAp > 2 m_{\chi}$, the parameter $Y$ is linked to the DM annihilation cross section via the formula~\cite{Gordan}:
\be
\sigma (\chi \bar \chi \to f \bar f) v  \propto \frac{8  \pi v^2 Y } { m_{\chi}^2}, 
\label{xthermal}
\ee
where $v$ is the relative velocity between the colliding DM particles. 
\section{The SHiP experiment}
\label{sec:SHiP_Experiment}
\begin{figure}[htb]
\centering
    \includegraphics[width=.99\textwidth]{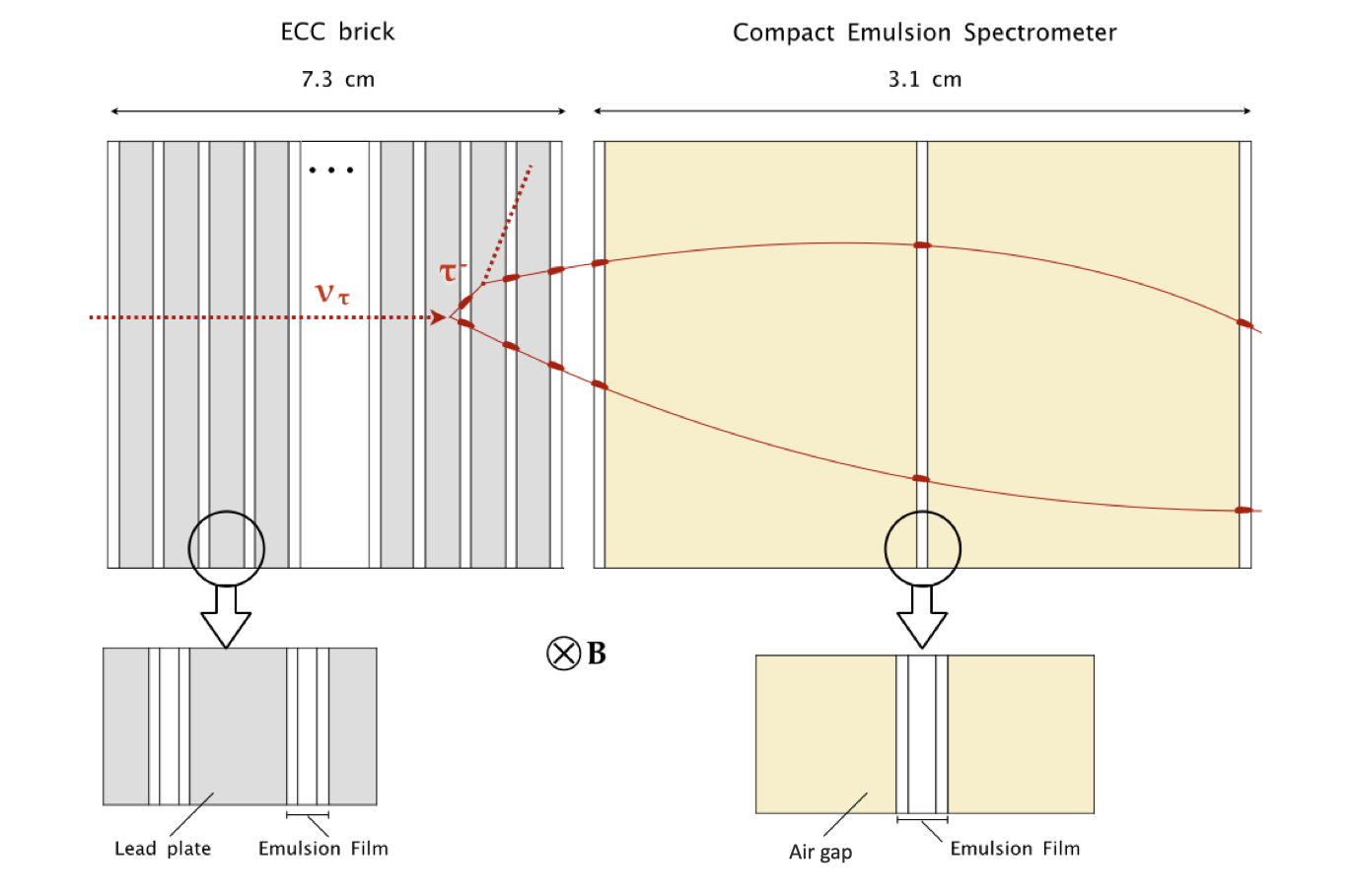}
    \caption{Schematic illustration of the basic unit of the Scattering Neutrino Detector and the ECC brick: on the left, emulsion films interleaved with lead plates; on the right, the Compact Emulsion Spectrometer.}\label{fig:brick}
\end{figure}

The Search for Hidden Particles (SHiP) experiment has been proposed as a general-purpose experiment~\cite{shiptp} at the CERN Super-Proton-Synchrotron (SPS), addressed to explore the high-intensity frontier for NP searches, thus complementing the LHC research program~\cite{shiptp}. It is particularly targeted at the observation of long-lived weakly interacting particles of mass below 10 GeV/c$^2$, foreseen in many Standard Model (SM) extensions. The use of a beam-dump facility~\cite{Ahdida_2019} will result in a copious flux of charmed hadrons, from which not only hidden sector particles originate~\cite{SHiP:2018xqw}, but also tau neutrinos and their corresponding anti-particles. Therefore, being also a neutrino factory, SHiP will perform a wide neutrino physics program, as well as a first direct observation of the tau anti-neutrino, which represents the last particle to be directly observed to complete the SM framework. The SHiP Scattering Neutrino Detector (SND) is an apparatus designed for LDM particles searches, since it exploits an optimised combination of a dense target and high-granularity scattering detector, being it based on nuclear emulsion technology.
\begin{figure}[htb]
\centering
    \includegraphics[width=1.\textwidth]{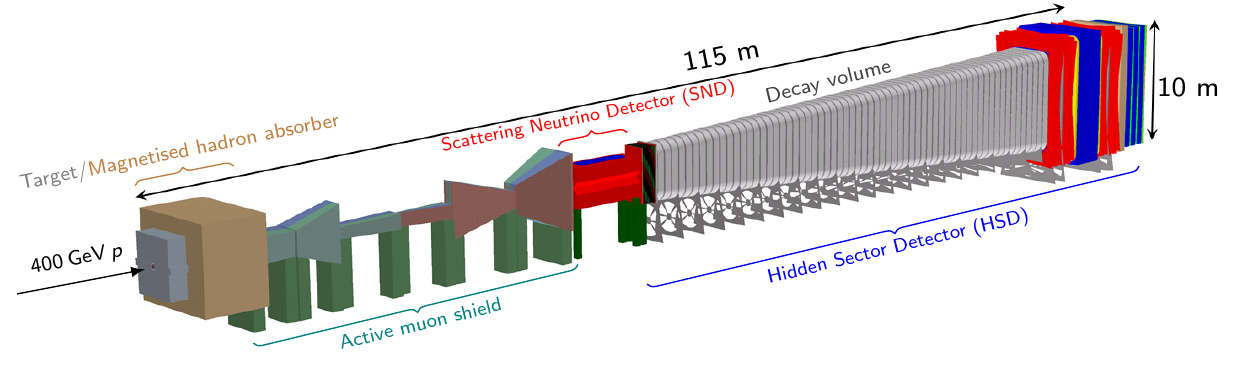}
    \caption{Overview of the SHiP experimental layout.} \label{facility}
\end{figure}
\begin{figure}[htb]
     \centering
     \begin{subfigure}[b]{0.64\textwidth}
         \centering
         \includegraphics[width=\textwidth]{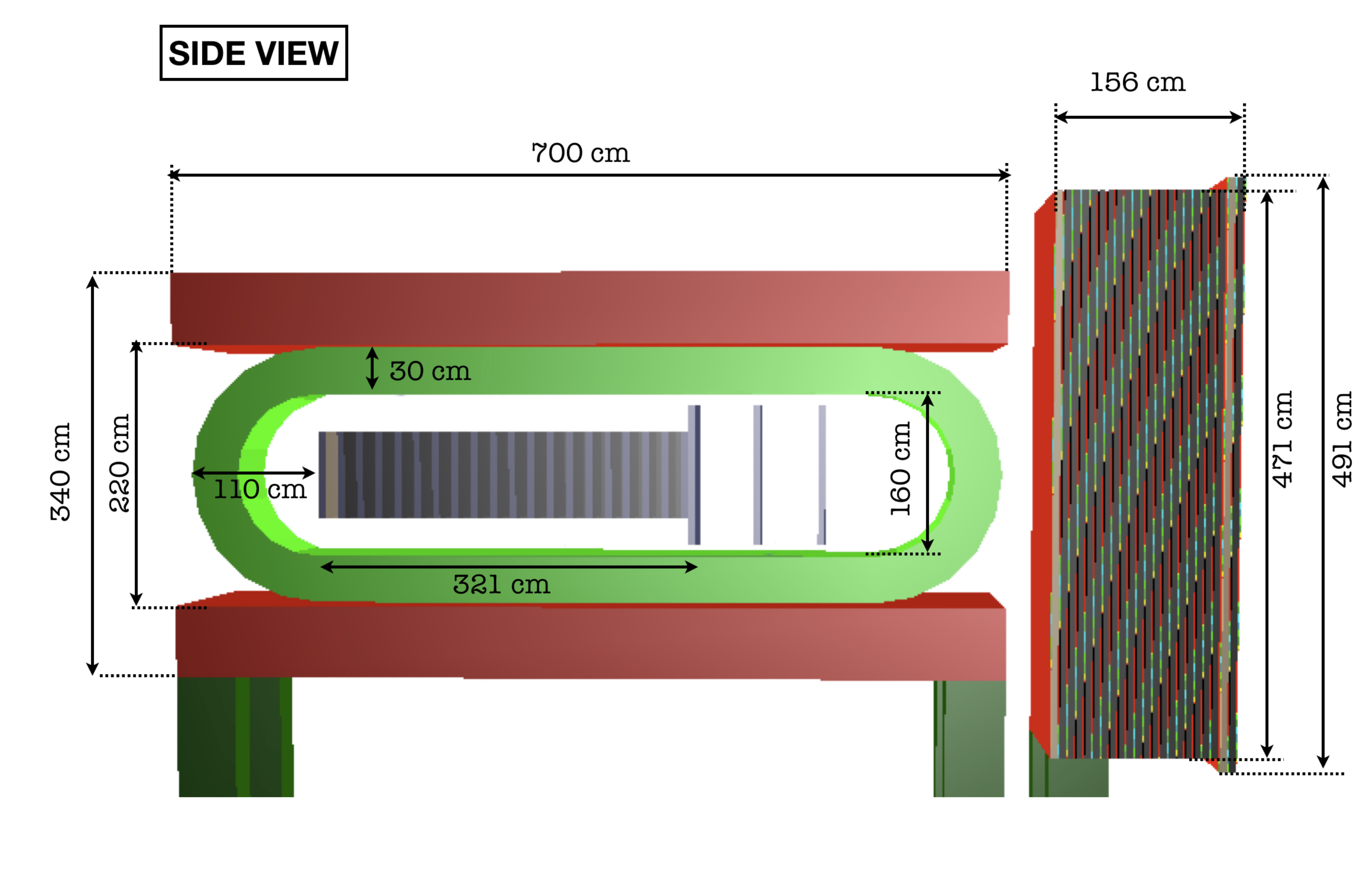}
         \caption{ }
         \label{subf:a}
     \end{subfigure}
     \hfill
     \begin{subfigure}[b]{0.35\textwidth}
         \centering
         \includegraphics[width=\textwidth]{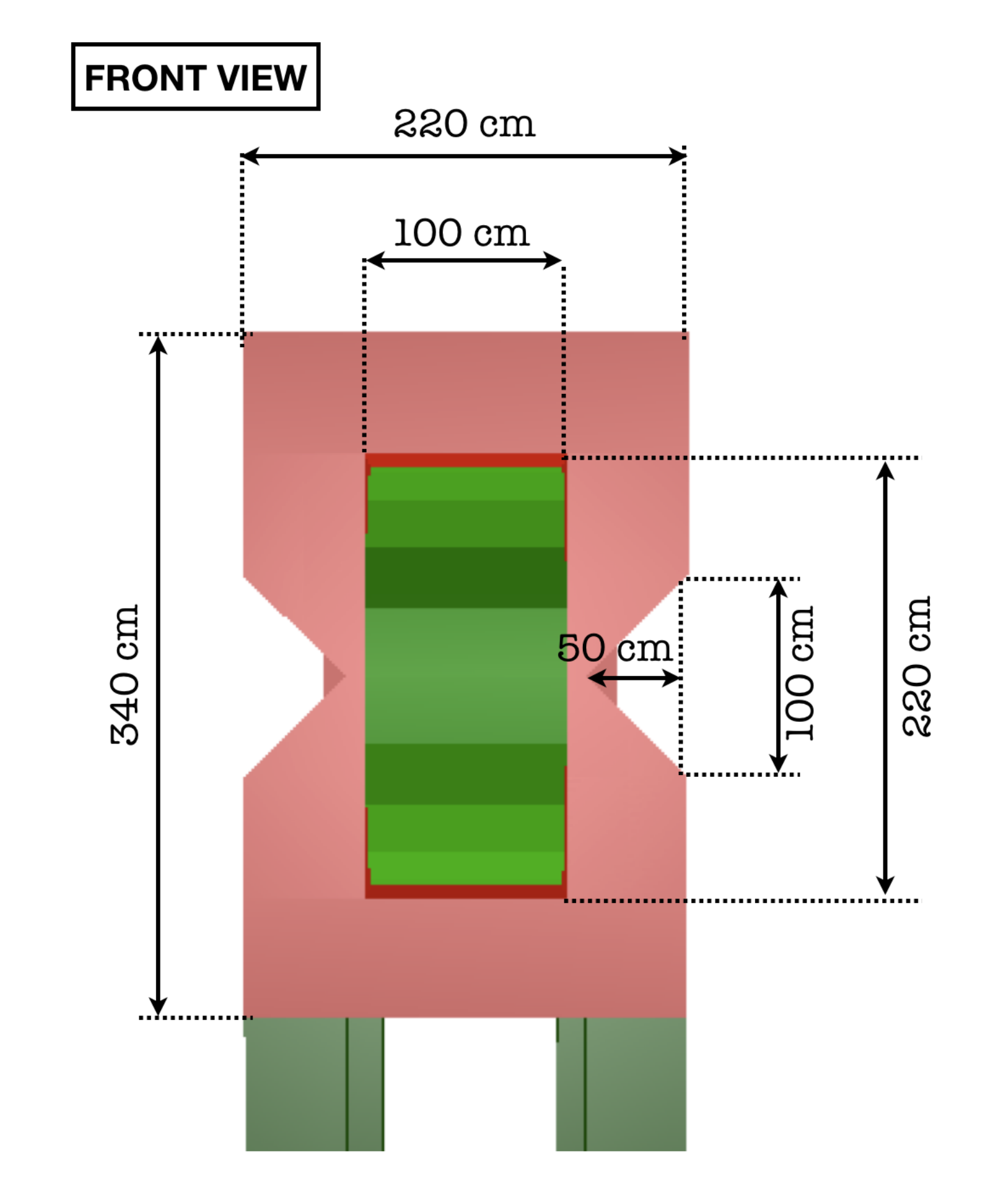}
         \caption{ }
         \label{subf:b}
     \end{subfigure}
     \caption{Side (a) and front (b) views of the Scattering Neutrino Detector layout adopted for this study, with a detail of the magnet (red) and of the coil (green).}
    \label{fig:snd}
\end{figure}
In Fig.~\ref{facility} a sketch of the experimental facility as currently implemented in the official simulation framework of the experiment FairShip~\cite{ref:FairShip} is shown. A synthetic overview of the simulated processes within FairShip and corresponding simulation software is reported in Tab.~\ref{tab:fairship}.\\
\indent
A 400 GeV/c proton beam will be delivered onto a thick heavy-metal hybrid target, specifically designed to maximise the charm production yield and thus hidden sector particles and tau neutrino yields. 
Over five years of SPS operations, a total of 2$\times$10$^{20}$ protons on target (\textit{p.o.t.}) collisions will be delivered, where each proton spill will have nominally 4$\times$10$^{13}$ \textit{p.o.t.}. A hadron stopper follows the beam-dump target, with the goal to absorb the SM particles produced in the beam interaction. In addition, a series of sweeping magnets, referred to as Muon Shield~\cite{Akmete_2017}, act as a deflecting device for the residual muons, further cleaning the flux of particles from leftover backgrounds to hidden sector particles and neutrino searches.\\
 \indent
The SND, shown in Fig.~\ref{fig:snd} in the setup adopted for this study, is located downstream of the muon sweeper. Placed in a magnetised region of $1.2$ Tesla in the horizontal direction and perpendicular to the beam axis~\cite{Ahdida_2020}, it consists of a $(90\times 75\times 321)\,$cm$^{3}$ high-granularity tracking device which exploits the Emulsion Cloud Chamber (ECC) technique developed by the OPERA experiment~\cite{Acquafredda_2009}, which was successfully used for tau neutrino detection~\cite{PhysRevLett.115.121802,opera1}. Each elementary unit of the modular detector, called brick (Fig.~\ref{fig:brick}), consists of alternating 56 lead plates of 1$\,$mm thickness, passive material to increase the interaction probability, and 57 nuclear emulsion films of 0.3$\,$mm thickness, acting as tracking detector with micro-metric accuracy. It is worth noting that the brick also functions as a high-granularity sampling calorimeter with more than five active layers for every radiation length $X_{0}$ over a total thickness of $10\,X_{0}$~\cite{4714e9793c854baa8f47e2269c4d04e0}. The ECC technology is also particularly efficient in the $e/\pi^0$ separation. The Compact Emulsion Spectrometer (CES), made up of a sequence of emulsion films and air gaps, is attached immediately downstream of the brick for electric charge measurement for particles not reaching the spectrometer. Despite the magnetic field, electron charge measurement is not possible due to early showering happening within the brick and the consequent information loss. The resulting weight of each ECC brick is approximately $8.3\,$kg, adding up to $\sim 8\,$tons for the whole SND. The bricks are then assembled to shape 19 walls of $\sim 50\,$units each, alternated with planes of electronic detector, called Target Tracker (TT), planes. For the time being, we consider the SciFi detector~\cite{ref:SciFi} as a feasible TT technological option. The TT additionally provide the time stamp of the event and help in linking the emulsion tracks to those reconstructed in the spectrometer and the muon system downstream of the SND. These features make the SND perfectly tailored for neutrino physics using all three flavours, as well as detection of light dark matter particles scattering off of electrons and nuclei of the SND.\\
An approximately 50~m long vacuum decay vessel is positioned downstream of the SND. The proposed facility is completed with a Hidden Sector Detector (HSD), equipped with calorimeters and muon detectors for the identification of long-lived Beyond Standard Model (BSM) particles.
\begin{table}[h]
    \centering
    \begin{tabular}{c  c}
         \hline
         Simulation & Software\\
         \hline
         \hline
         SHiP detector: geometry and transport & \texttt{GEANT4}~\cite{AGOSTINELLI2003250} \\
         Proton on target collisions & \texttt{PYTHIA v8.2}~\cite{Sjostrand:2014zea}\\
         Heavy flavour cascade production & \texttt{PYTHIA v6.4}~\cite{Sj_strand_2006}\\
         Neutrino interactions & \texttt{GENIE}~\cite{ANDREOPOULOS201087}\\
         \hline
    \end{tabular}
    \caption{Details of the different steps of the simulation process within the FairShip framework and corresponding employed software.}
    \label{tab:fairship}
\end{table}

\section{Light dark matter production and detection}
\label{sec:LDM_prod_detect}
\begin{figure}[htb]
\centering
    \includegraphics[width=0.5\textwidth]{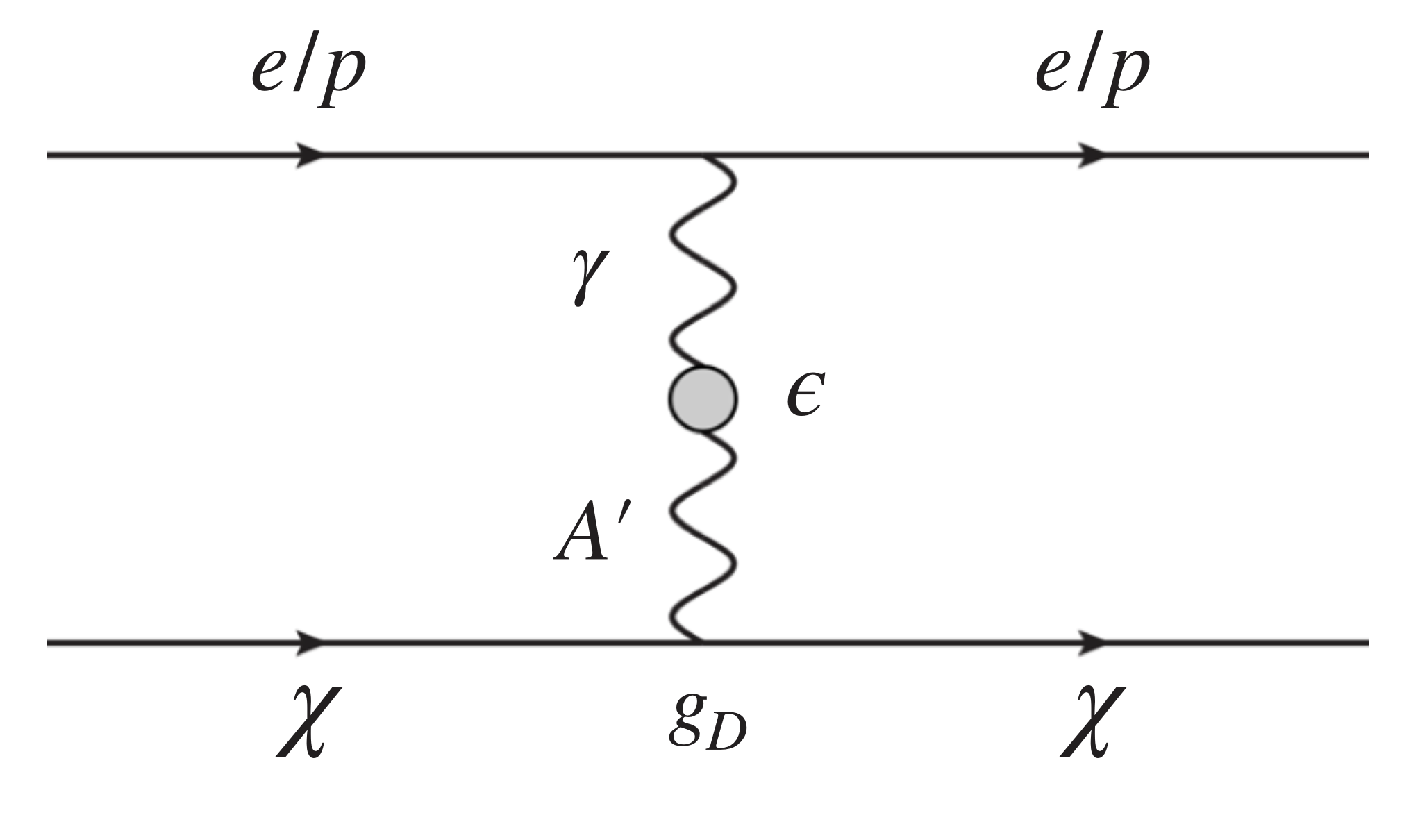}
    \caption{Light dark matter interaction processes which can be probed by the SHiP experiment within the Scattering Neutrino Detector, \textit{i.e.} elastic scattering off electrons $(\chi\,e^{-}\to\chi\,e^{-})$ and off protons $(\chi\, p\to\chi\, p)$.} \label{DMscattering}
\end{figure}

At a proton beam dump, DP can be abundantly produced in several channels:
\begin{enumerate}
    \item \textit{Light meson decay}: proton collisions on a heavy target result in a copious flux of outgoing mesons. Hence, DP may be produced in radiative decays of neutral mesons, whereas a final state photon converts into a DP. 
    The production cross-section is proportional to $\epsilon^{2}$ and the relevant contributions come from the lightest mesons, because of decay modes with photons with relatively high branching ratio: $\pi^0$, $\eta$, $\omega$~\cite{Batell:2009di}.
    \item \textit{Proton bremsstrahlung}: being a charged particle surrounded by its own electromagnetic field, the proton radiates low-frequency and/or quasi-collinear photons with high probability when it undergoes a scattering process. 
    Vector states like DP mediators can then be generated via radiative process $p\,A\to p\,A\,\Ap$~\cite{Blumlein:2013cua} in proton interactions with the target nuclei.
    \item \textit{Direct perturbative QCD production}: it corresponds to the dominant production mechanism for higher masses ($\mAp\gtrsim 1\,$GeV). At the lowest order in the strong interaction, DP are produced through the quark-antiquark annihilation process $q\bar{q}\to\Ap$ ~\cite{Batell:2009di}. At higher orders, one can also have the associated production with a jet, according to the the quark-gluon scattering process $qg\to q\Ap$, and with multiple jets.
\end{enumerate}
\indent
In addition, secondary leptons produced in the dump can contribute to the flux of photons, and thereby of DPs, by
different re-scattering processes occurring within the target. 
Such lepton-induced processes are usually sub-dominant at a proton beam dump. However, they are not completely negligible, as nicely shown in a the dedicated analysis~\cite{Celentano:2020vtu}, and might be relevant in scenarios in which the New Physics does not couple with colour particles.  
We do not include them in this work. Therefore, our estimates should be considered conservative in this regard. 
The minimal DP model can be probed by the SHiP experiment through the direct detection of LDM elastic scattering process off of the electrons and nuclei of the SND (Fig.~\ref{DMscattering})
For the majority of the events $\chi\,e^{-}\to \chi\, e^{-}$, the scattered electron is sufficiently energetic to generate an electromagnetic shower within the brick. Given that the ECC device acts as a high-granularity sampling calorimeter, it is thus possible to reconstruct the electron and measure its energy. Furthermore any activity in the proximity of the primary vertex can be spotted down to 100 MeV or below, thanks to the micrometrical position resolution of the nuclear emulsion device and the high sampling rate. These features translate into capability to accurately identify and tackle background events to LDM searches, as further described in Sec.~\ref{sec:Bkg}. As a consequence, LDM scattering events can be distinguished from a large neutrino-induced background.\\
\indent
An estimate of the order of magnitude of the expected yield of LDM interactions at SHiP can be determined as follows. The number of LDM-electron scattering events in the SND detector is given by the standard formula
\be
\label{eq:ldm1}
\mathcal{N}_{\mathrm{LDM}} =\sigma(\chi\,e^-\to\chi\,e^-)\cdot\frac{\phi}{A_{\mathrm{\tiny SND}}}\cdot N_{e^{-}}\,,
\ee
where $N_{e^{-}}$ is the numbers of scattering centres, $i.e.$ the electrons in the detector, $\phi$ is the flux of incident LDM particles and $A_{\mathrm{\tiny SND}}$ represents the transverse area in $(x,\,y)$ of the SND. The elastic LDM-electron scattering cross section is roughly given by
\be
\label{eq:ldm2}
\sigma(\chi\,e^-\to\chi\,e^-)\simeq \frac{4\,\pi\,\alpha\,\alpha_D\,\epsilon^2}{m^2_{A^{'}}}\,.
\ee
The flux $\phi$ mainly depends on the specific value of the DP mass which in turn 
determines the relative importance of the different production mechanisms. 
For example, for $\mAp\ll m_{\pi}$, LDM production in the beam dump is dominated by pion decays. In this case and under the assumption that all the primary proton impinging on the target will eventually interact in the beam dump, $\phi$ can be written as  
\be
\label{eq:ldm3}
\phi\,\simeq\, 2\cdot N_{p.o.t}\cdot\lambda_{\pi^0}\cdot\epsilon^2\cdot \mathcal{A_{\mathrm{\tiny geo}}}\,.
\ee
In Eq.~\eqref{eq:ldm3}, $N_{p.o.t.}$ is the total number of $p.o.t.$ delivered in the five years of data-taking; $\lambda_{\pi^0}$ denotes the multiplicity of $\pi^{0}$s per $p.o.t.$; $\mathcal{A}_{\mathrm{\tiny geo}}$ embeds the geometrical acceptance of the SND to LDM interaction vertices, corresponding to an angular coverage $|(\theta_{x},\,\theta_{y})|\leq(12,\,10)$ mrad from the proton beam dump. If considering an average value of $\lambda_{\pi^0}\sim 6$ as provided by the simulation of prompt proton-nucleon collisions with \texttt{Pythia}\footnote{We use \texttt{Pythia}(v8.230) and simulate events under the flag \textit{SoftQCD:Inelastic.}}~\cite{Sjostrand:2014zea}, a geometrical acceptance $A_{geo}\sim30{}\%$ and if assuming a coupling close to the current experimental constraints $\epsilon~\sim~5\times~10^{-5}$ for a 10 MeV-DP and $\alpha_D \sim 0.1$, the expected number of LDM candidates foreseen in SHiP is $\sim~1.3\times 10^4$. 

We used MadDump~\cite{Buonocore2019} as the principal tool for the simulation of signal events. Its general philosophy and all the technical details are outlined in Ref.~\cite{Buonocore2019}. 
We generate the event samples at the particle level and apply the selection criteria on the recoil electrons without taking into account other detector effects besides the geometrical acceptance. 
This is consistent with what has been done in the estimate of the background event rate. Since the target length is way larger than the proton interaction length in the material, we assume all of them to interact within the beam dump.
In the following, we give further details for each production mechanism.
\subsection{Meson decay}
\begin{figure}[t]
    \centering
    \includegraphics[width=0.5\textwidth]{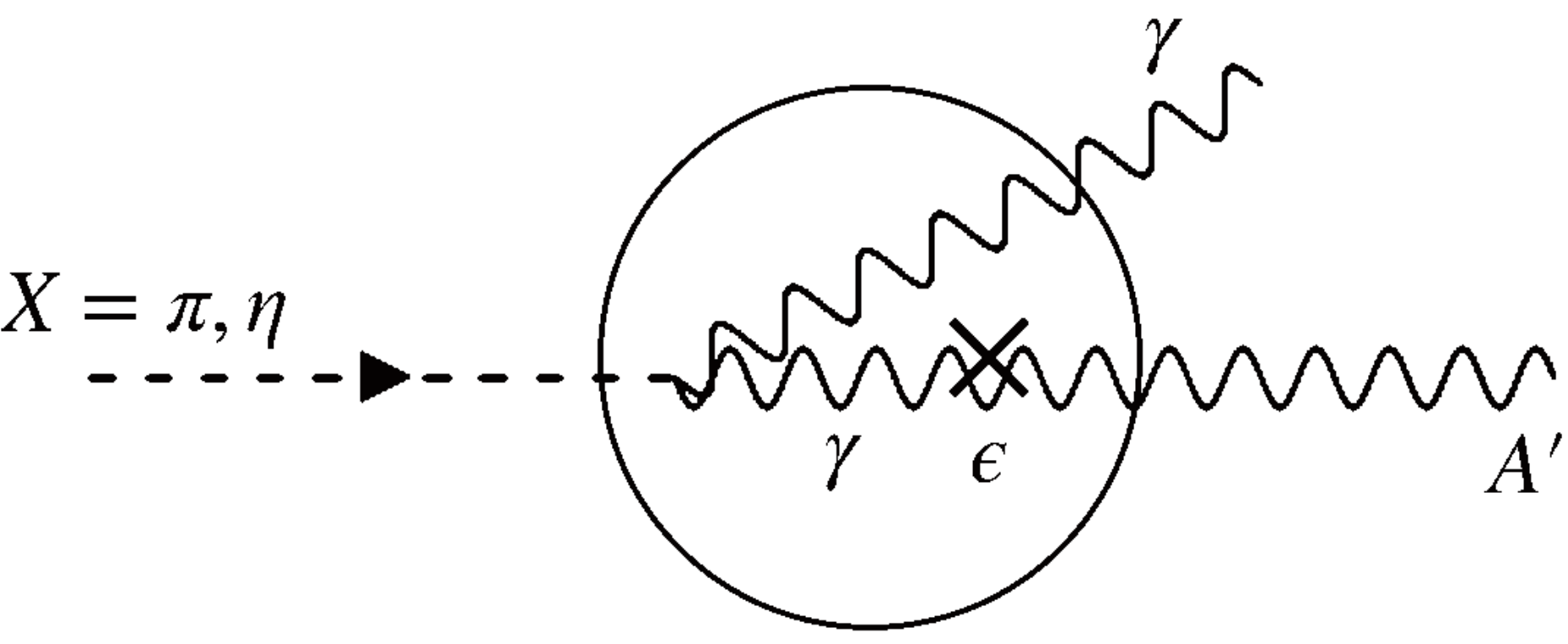}
    \caption{ Effective vertex for the decay process $X\to\gamma \Ap,\, X=\pi^0,\eta$.  }
    \label{fig:eft-vertex}
\end{figure}
The relevant parameter space within the reach of the SHiP SND corresponds 
to $ \mAp > 2 m_{\chi}$ and $ g_D \gg \epsilon e$.
Indeed, in this scenario, the DP decays almost entirely into DM after travelling 
a very short distance, maximising the DM flux reaching the 
SHiP SND. The decay rate for light mesons decaying into dark photons is then dominated 
by the formation of an on-shell dark photon which decays promptly into dark matter, 
$BR(\Ap\to\chi\bar{\chi})\simeq 1$. The production process is then well described by an effective Lagrangian with mesons as dynamical degrees of freedom leading to interaction vertices like $X\gamma \Ap$ ($X=\pi^0,\eta$, see Fig.~\ref{fig:eft-vertex}) and $\omega\pi^0\Ap$. 
The corresponding branching ratios scale with $\epsilon^2$ and are given by:
\begin{align}
\label{eq:decayrates}
&\frac{\text{BR}(\pi^0 , \eta, \eta^\prime \to \gamma \Ap)}{\text{BR}(\pi^0, \eta, \eta^\prime \to \gamma \gamma)} \simeq 2 \, \epsilon^2 \, \Big( 1 - \frac{\mAp^2}{m_{\pi^0, \eta, \eta^\prime}^2}\Big)^3
\\
& \frac{\text{BR}(\omega \to \pi^0 \Ap)}{\text{BR}(\omega \to \pi^0 \gamma)} \simeq \epsilon^2 \, \left( m_{\omega}^2 - m_\pi^2\right)^{-3} \Big[ (\mAp^2 - (m_\pi + m_\omega)^{2}) (\mAp^2 - (m_\pi - m_\omega)^2) \Big]^{3/2}.
\end{align}
An interested reader can find useful insights about the formulas above in~\cite{Kahn:2014sra,Gardner:2015wea,deNiverville:2011it}.
The full simulation process is performed in three steps: 
\begin{itemize}
    \item[i.] production of the input meson fluxes originating from the incoming protons impinging and interacting within the target (beam dump)
    \item[ii.] generation of DM fluxes from the BSM meson decays in the relevant DM mass range   
    \item[iii.] generation of the corresponding $\mathrm{DM}-e^-$ scattering events within the detector acceptance and the selection criteria. 
\end{itemize}

MadDump provides a unified framework to handle the last two steps, in which all the new physics content is involved. 
The main source of uncertainties comes from the meson fluxes.  
Indeed, the description of the proton-nucleus interactions is highly non-trivial and experimental data are available only for a limited collection of energies and nuclear targets. 
Phenomenological and data-driven parametrisations for the distributions of the light mesons have been proposed in the literature \cite{Bonesini:2001iz}.
An alternative strategy is provided by Monte Carlo programs like \texttt{Pythia}~\cite{Sjostrand:2014zea}. Recently, \texttt{Pythia(8)} results have been compared with existing experimental data for the inclusive production of $\pi^0$ and $\eta$ mesons in $pp$ collisions~\cite{Dobrich:2019dxc}. A fairly good agreement has been found for mesons with high momentum and within middle-high rapidity range (where the Feynman variable $0.025<x_{\mathrm{F}}<0.3$), which represent the bulk of our events in acceptance. 

Furthermore, secondary interactions of hadrons in the beam-dump target may affect the particle multiplicities, which in turn may increase the LDM yields and impact the sensitivity reach of the experiment. 
It is thus important to estimate the so-called cascade effects \cite{cascade}.
As the main input for the lightest mesons ($\pi^0,\eta$) we use the fluxes generated with \texttt{GEANT4(v10.3.2)} within the FairShip software framework, which takes into account the secondary interactions adapting what has been used in Ref.~\cite{SHiP:2020hyy}. We also consider samples of mesons from primary proton-nucleon interactions generated with \texttt{Pythia}, as a reference to assess the impact of the cascade. For the $\omega$, we rely on the \texttt{Pythia} samples only.

\begin{figure}[h]
    \centering
    \includegraphics[width=0.49\textwidth]{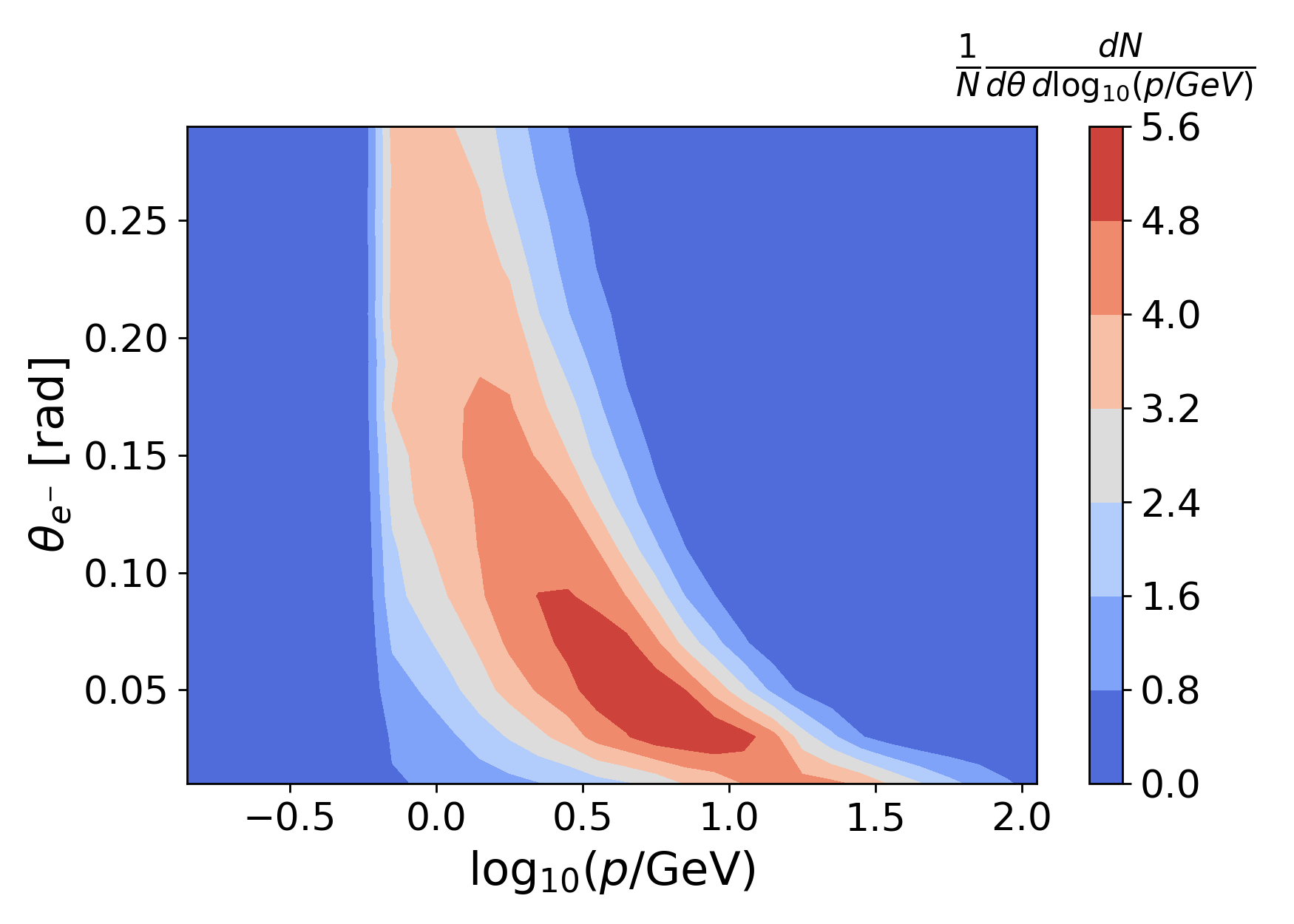}
    \hfill
    \includegraphics[width=0.49\textwidth]{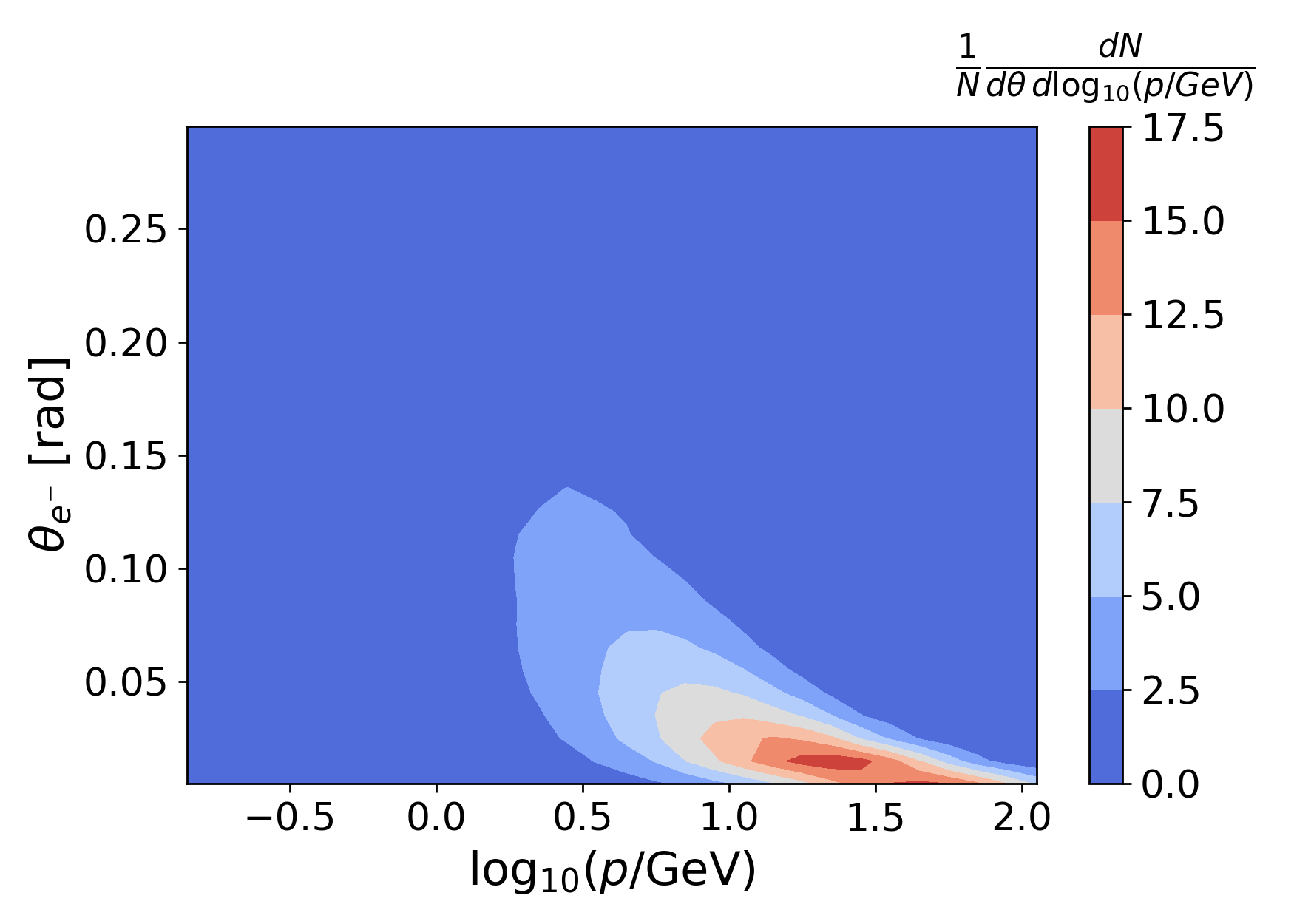}
    \caption{2D contour plot of the momentum ($p$) and the 
    production angle ($\theta$) correlation for $\pi^0$s produced in the collisions of $400\,$GeV protons hitting the SHiP beam-dump target.
    Left: simulation with FairShip including $\pi^0$ production in the interactions of secondary hadrons with the target nuclei. Right: simulation of the prompt proton-nucleon $\pi^0$ production with \texttt{Pythia}.}
    \label{fig:cmp-pion-spectrum}
\end{figure}

\begin{figure}[h]
    \centering
    \includegraphics[width=0.49\textwidth]{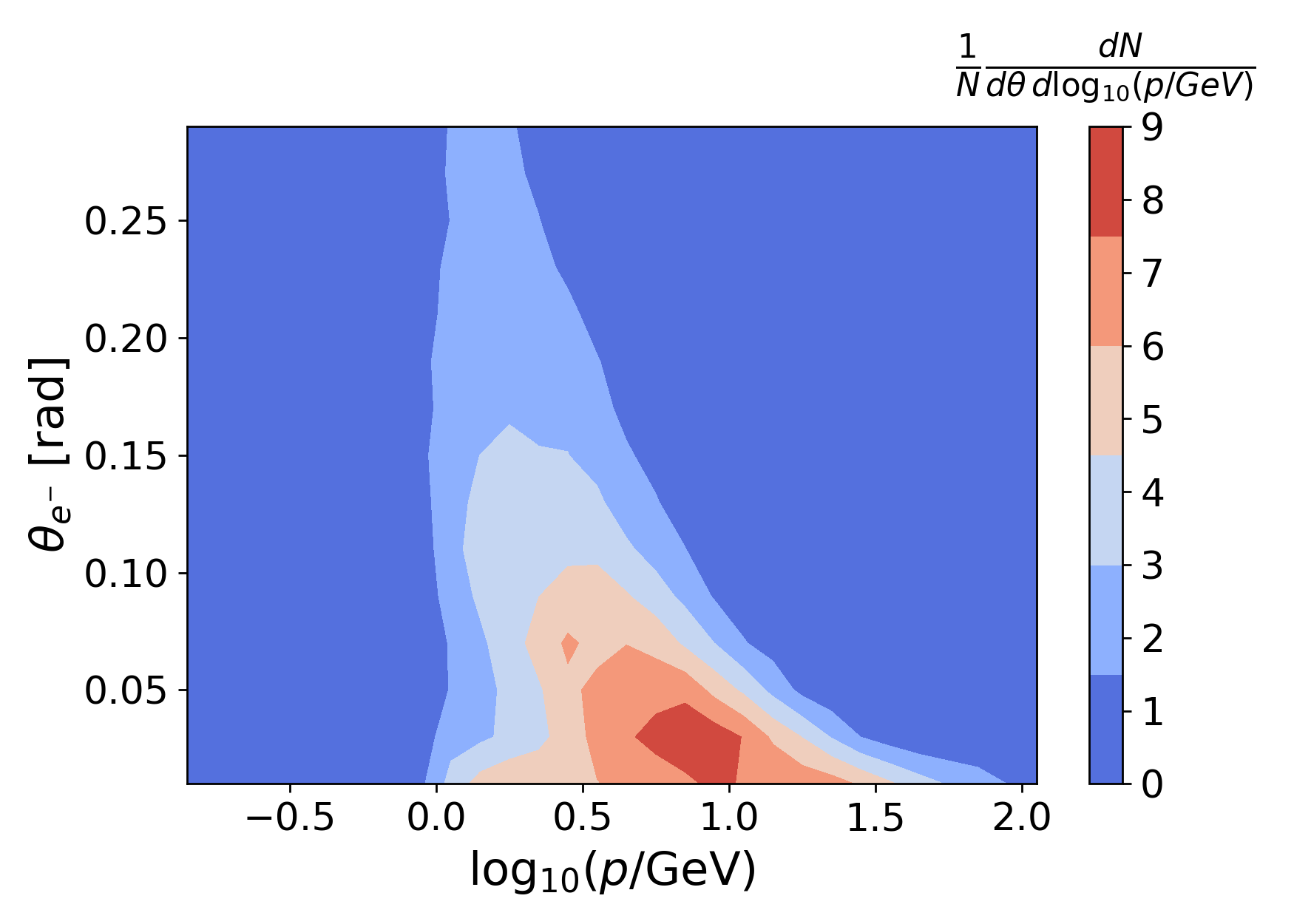}
    \hfill
    \includegraphics[width=0.49\textwidth]{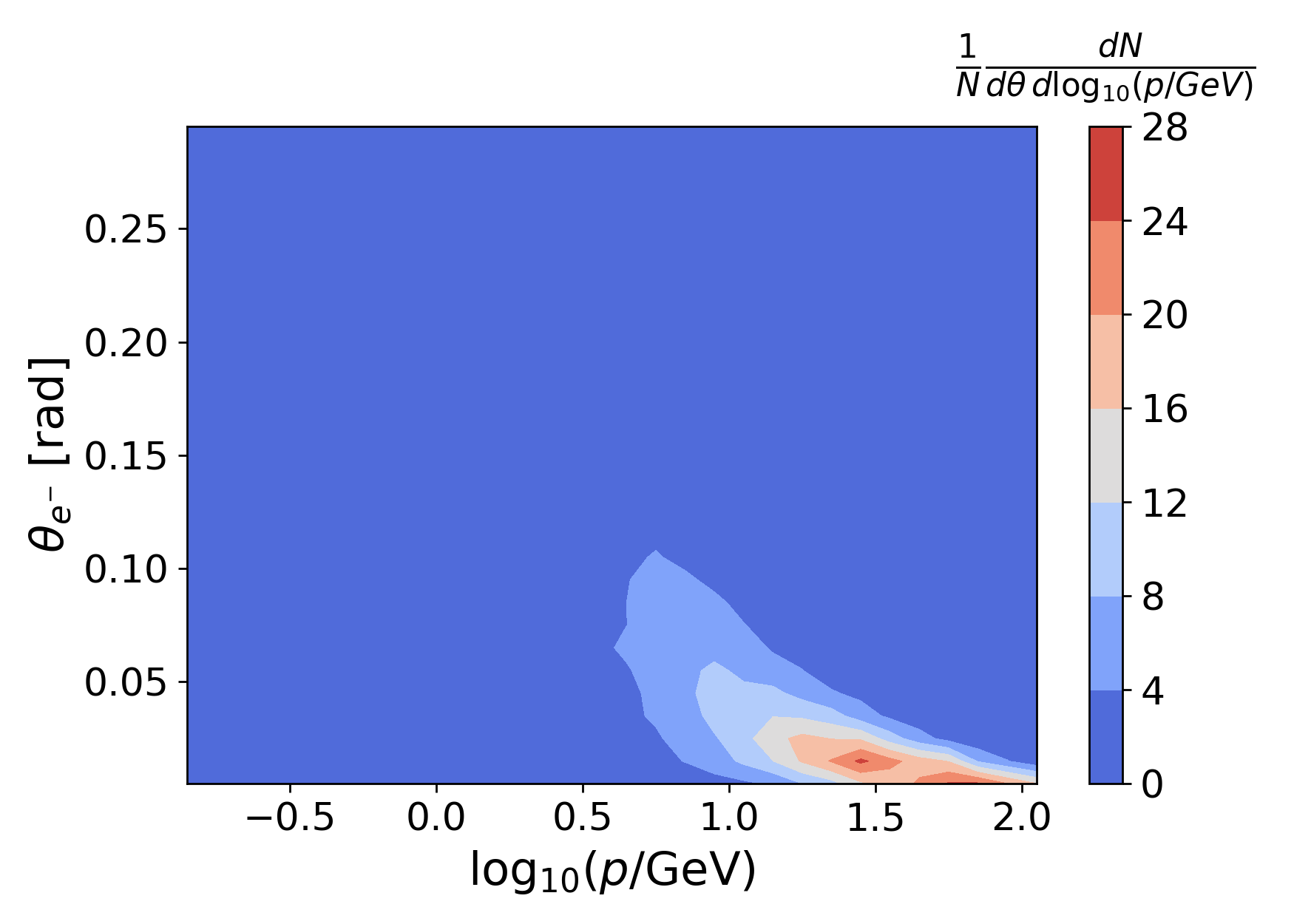}
    \caption{2D contour plot of the momentum ($p$) and the production angle ($\theta$) correlation for the $\eta$s produced in the collisions of $400\,$GeV protons hitting the SHiP beam-dump target.
    Left: simulation with FairShip including $\eta$ production in the interactions of secondary hadrons with the target nuclei. Right: simulation of the prompt proton-nucleon $\eta$ production with \texttt{Pythia}.}
    \label{fig:cmp-eta-spectrum}
\end{figure}

In Tables~\ref{tab:cmp-meson-number},~\ref{tab:cmp_pion} and~\ref{tab:cmp_eta}, we report a selection of results for $\pi^0$ and $\eta$, comparing the FairShip and \texttt{Pythia} samples. An important parameter in the FairShip simulation is the energy cut-off $E_\text{cut}$ applied to the particles produced at each interaction vertex: particles with energy less than $E_\text{cut}$ are removed from the list of those considered for new interactions within the target. We report the result for $E_\text{cut} > 500\,$MeV. Primary proton-proton interactions, as generated with \texttt{Pythia}, give an average particle multiplicity per \textit{p.o.t.} of ${N_{\pi^0}}/{p.o.t.}=6$ and ${N_{\eta}}/{p.o.t.}=0.8$,  for $\pi^0$ and $\eta$ respectively. From the samples of mesons generated with FairShip, we get ${N_{\pi^0}}/{p.o.t.}=42$ and ${N_{\eta}}/{p.o.t.}=5.5$. Therefore, we observe that secondary interactions occurring within the beam-dump target greatly increase the particle multiplicities and, in turn, lead to a rise of the DM yield by the same amount. 
However, this does not translate directly into an enhancement of the signal yield in the SND. Indeed, in order to produce a detectable scattering event one should take into account
\begin{itemize}
    \item the geometrical acceptance,
    \item the path travelled within the detector,
    \item the cross section for the scattering process.
\end{itemize}
We consider separately the effect due to the geometrical acceptance, defining an effective number of mesons per \textit{p.o.t.} ${N^{\mathrm{eff}}_X}/{p.o.t.}$ as the average number of mesons of species $X$ per \textit{p.o.t.} which produce a DM particle impinging on the detector surface. 
For different $\mAp$ values, we report in Tables~\ref{tab:cmp_pion} and~\ref{tab:cmp_eta} our estimate of ${N^{\mathrm{eff}}_{\pi^0}}/{p.o.t.}$ and ${N^{\mathrm{eff}}_{\eta}}/{p.o.t.}$ as estimated with \texttt{Pythia} and FairShip. The comparison shows that the increase due to the cascade is around $50-70\%$.
The explanation is that the secondary particles mainly populate the soft part of the spectrum, as it is clearly shown in the left panels of Fig.~\ref{fig:cmp-pion-spectrum} and Fig.~\ref{fig:cmp-eta-spectrum} which have to be compared with the corresponding right panels describing the spectrum from prompt yields. 
Moreover, the cross section for the elastic LDM-$e^-$ scattering grows with the energy of the incoming dark-matter particle before saturating to a constant behaviour~\cite{Buonocore:2019esg}. Hence, we expect that scattering events initiated by LDM particles produced in secondary interactions, being softer, will be less probable. This is clearly demonstrated by the last two columns in Tables~\ref{tab:cmp_pion} and~\ref{tab:cmp_eta} in which we report the final signal yields $N_s$ (corresponding to the benchmark point $\alpha_D= 0.1$ and $\mAp= 3 m_\chi$ and $\epsilon=10^{-4}$) due to FairShip and \texttt{Pythia} samples respectively. From the comparison, we see that the impact of the secondary interactions is reduced to that given by the geometrical acceptance only. 
In conclusion, our finding is that for $\pi^0$, the cascade modestly affects $(\sim 15-40\%)$ the signal event yields within the detector volume, while for $\eta$ it is negligible. 

\begin{table}[h]
    \centering
    \begin{tabular}{c|c|c}
    \hline
        meson & ${N_{\pi^0}}/{p.o.t.}$ & ${N_{\pi^0}}/{p.o.t.}$    \\
              & FairShip & \texttt{Pythia} \\ 
         \hline
         $\pi^0$ & $42$ & $6$  \\
         $\eta$  & $5.5$ & $0.8$ \\
         \hline
    \end{tabular}
    \caption{Average particle multiplicities per \textit{p.o.t.} in $400\,$GeV proton collisions as estimated with FairShip, applying a cut-off $E_\text{cut}>500\,$MeV on secondary particles, and with \texttt{Pythia}, for primary interactions only. }
    \label{tab:cmp-meson-number}
\end{table}

\begin{table}[h]
    \centering
    \begin{tabular}{c|c|c|c|c}
    \hline
    $\mAp\,$(MeV) &    ${N_{\pi^0}^{\mathrm{eff}}}/{p.o.t.}$ & ${N_{\pi^0}^{\mathrm{eff}}}/{p.o.t.}$  & ${N_s}$ & ${N_s}$   \\
            &   (FairShip) &    (\texttt{Pythia}) &   (FairShip) &    (\texttt{Pythia})\\
    \hline
    $10$  &   $1.2$	&   $0.80$    &   $1.7\cdot 10^{4}$   &    $1.3\cdot 10^{4}$\\ 	
    $30$  &   $1.1$	&   $0.72$    &   $8.6\cdot 10^{3}$   &   $7.3\cdot 10^{3}$  \\	
    $60$  &   $0.70$	&   $0.46$    &   $2.0\cdot 10^{3}$   &   $1.8\cdot 10^{3}$  \\	
    $90$  &   $0.24$	&   $0.15$    &   $3.1\cdot 10^{2}$   &   $2.5\cdot 10^{2}$   \\	
    $120$ &   $0.013$	&   $0.0083$    &   $7.4$   &   $6.7$  \\	
    \hline
    \end{tabular}
    \caption{Comparison between $\pi^0$ samples generated using FairShip (with an energy cut of $E_{\rm cut} > 500\,$MeV in secondary vertices) and \texttt{Pythia}.  ${N_{\pi^0}^{\mathrm{eff}}}/{p.o.t.}$ is the effective number of $\pi^0$s per \text{p.o.t.} producing LDM particles within the geometrical acceptance.
    $N_s$ is the signal yield for the benchmark point $\alpha_{D}= 0.1$ and $\mAp= 3 m_\chi$ and $\epsilon=10^{-4}$ corresponding to $2\cdot 10^{20}p.o.t.$
    }
    \label{tab:cmp_pion}
    \end{table}

\begin{table}[h]
    \centering
   \begin{tabular}{c|c|c|c|c}
   \hline
    $\mAp\,$(MeV) &    ${N_{\eta}^{\mathrm{eff}}}/{p.o.t.}$ & ${N_{\eta}^{\mathrm{eff}}}/{p.o.t.}$ & ${N_s}$ & ${N_s}$  \\
            &   (FairShip) &    (\texttt{Pythia}) &   (FairShip) &    (\texttt{Pythia}) \\
    \hline
    $10$  &   $0.15$	&   $0.10$    &   $1.1\cdot 10^{3}$   &    $8.1\cdot 10^{2}$\\ 	
    $130$ &   $0.13$	&   $0.092$    &   $25$   &    $24$\\
    $250$ &   $0.099$	&   $0.059$    &   $1.6$   &    $1.5$\\
    $370$ &   $0.033$	&   $0.020$    &   $1.16\cdot 10^{-1}$  &    $1.15\cdot 10^{-1}$\\	
    $520$ &   $0.00020$	&   $0.00012$    &   $1.9\cdot 10^{-4}$   &   $1.8\cdot 10^{-4}$   \\
    \hline
    \end{tabular}
    \caption{Comparison between $\eta$ samples generated using FairShip (with an energy cut of $E_{\rm cut} > 500\,$MeV in secondary vertices) and \texttt{Pythia}.  ${N_{\pi^0}^{\mathrm{eff}}}/{p.o.t.}$ is the effective number of $\eta$s per \text{p.o.t.} which give raise to LDM particles within the geometrical acceptance. 
    $N_s$ is the LDM yield for the benchmark point $\alpha_D= 0.1$ and $\mAp= 3 m_\chi$ and $\epsilon=10^{-4}$
    corresponding to $2\cdot 10^{20}p.o.t.$
    }
    \label{tab:cmp_eta}
\end{table}

\clearpage
\subsection{Proton Bremsstrahlung}
In the mass range $500\,$MeV  $\lesssim \mAp \lesssim 1\,$GeV, the production of $\Ap$ is dominated 
by the proton bremsstrahlung mechanism. 
The photon emission cross section is indeed enlarged in the collinear direction so that a sizeable
fraction of $\Ap$ is produced within the geometrical acceptance for an on-axis detector as the SND ($\sim 20\%$).
In this limit, the process can be described by a generalisation of the Fermi-Williams-Weizsäcker method~\cite{Fermi:1924tc,Williams:1934ad,vonWeizsacker:1934nji}, based on the assumption that the $p-N$ scattering is dominated by the exchange in the $1^{--}$ channel.
We extend MadDump include the bremsstrahlung from the primary protons. Following Refs.~\cite{Blumlein:2013cua,PhysRevD.95.035006},
we parametrise the four-momentum vector of the emitted $\Ap$ as $p_{\Ap} = (E_{\Ap}, p_{\mathrm{T}} \cos(\phi), p_{\mathrm{T}} \sin(\phi), zP)$,  with $E_{\Ap}=zP + (p_{\mathrm{T}}^2+\mAp^2)/(2zP)$, where $P$ is the momentum of the incident proton, $z$ is the fraction of the proton momentum carried by the outgoing $\Ap$, $p_{\mathrm{T}}$ is the momentum perpendicular to the beam momentum and $\phi$ is the azimuthal angle. We generate unweighted $\Ap$ events according to the differential production rate
\begin{equation}
\label{eq:bremrate}
	\frac{d^2 N_{\Ap}}{dz dp^2_{\mathrm{T}}} = \frac{\sigma_{pA}(s^\prime)}{\sigma_{pA}(s)} F^2_{1,p}(\mAp^2) w_{ba}(z,p^{2}_{\mathrm{T}}),
\end{equation}
where $s^\prime=2m_p(E_p - E_{\Ap})$, $s=2m_p E_p$ and the photon splitting function is
\begin{equation}    
\begin{aligned}
\label{eq:wba}
	w_{ba}(z,p^2_{\mathrm{T}}) =& \frac{\epsilon^2\alpha}{2\pi H}\Bigg[ \frac{1+(1-z)^2}{z} -2z(1-z)\left(\frac{2m_p^2+\mAp^2}{H}-z^2\frac{2m_p^4}{H^2}\right) \nonumber\\
	 &\qquad\qquad+2z(1-z)(1+(1-z)^2)\frac{m_p^2 \mAp^2}{H^2}+2z(1-z)^2\frac{\mAp^4}{H^2}\Bigg],
\end{aligned}
\end{equation}
with $H = p^2_{\mathrm{T}} + (1-z)\mAp^2+z^2 m_p^2$.
In the above formula, $F_{1,p}$ is the time-like proton form-factor, as provided by the parameterisation in Ref.~\cite{Faessler:2009tn}. It effectively incorporates off-shell mixing with vector mesons such as $\rho$ and $\omega$, corresponding to a resonance effect around $\mAp\sim 770\,$MeV~\cite{Morrissey:2014yma}. In Ref.~\cite{PhysRevD.95.035006}, 
the authors compare the description of the peak region by adopting the time-like proton form factors and by adding by hand
the vector mixing within an on-shell computation, finding small deviations in the peak region. Assessing the size of this uncertainty is beyond the scope of this work. 

The next steps of the simulation, namely the decay $\Ap \to \chi \bar{\chi}$
and the $\chi-e^-$ scattering in the SND, are handled by standard MadDump functions. The whole process has been
integrated into the new MadDump mode \texttt{bremsstrahlung-interaction}. 

The normalisation of the flux of the original $\Ap$ is given by the integral of the differential production rate eq.~\eqref{eq:bremrate} 
in the validity range of the equivalent photon approximation, given by the kinematical conditions 
\begin{equation}
\label{eq:bremcond}
E_p,E_{\Ap},E_p-E_{\Ap} \gg m_p, \mAp, |p_{\mathrm{T}}|.
\end{equation}
Following Refs.~\cite{Blumlein:2013cua,PhysRevD.95.035006,Gorbunov:2014wqa}, we adopt 
$z\in[0.1,0.9]$. For a relatively high energy experiment such as SHiP, the minimum DP energy $E_P$ corresponds then to $\sim40\,${}GeV and we can set 
its maximum transverse momentum $p_T$ to 4 GeV, i.e. an order of magnitude less.
We expect electron bremsstrahlung to be sub-dominant as discussed for example in \cite{Gorbunov:2014wqa,Berlin:2018pwi}. As for the cascade effects, extra dark photons may arise from the
bremsstrahlung of secondary charged hadrons. Similarly to what happens in the case of mesons, the picture is complicated by the impact of the geometrical acceptance and the convolution
with the scattering cross section. For the case the proton undergoes a chain of elastic proton-nucleon collisions, so that it retains all of its initial energy, 
we can make a rough estimate by means of the following simplified calculation. 
Let $p_{\rm el}$ be the probability that the incoming proton undergoes an elastic scattering interaction with a nucleon in the target and $p_{\rm brem}$ the probability of a dark photon produced in the proton bremsstrahlung. 
Under the assumption that $p_{\rm brem}$ does not depend much on the number of previous elastic collisions, the probability that a dark photon is produced in this chain is 
\begin{equation}
    p = p_{\rm brem}\left(1+p_{\rm el}\times p_{\rm el}+p_{\rm el}\times p_{\rm el}\times p_{\rm el}+\dots \right) = p_{\rm brem}\sum_{n=0}^{\infty}p_{\rm el}^n = p_{\rm brem} \frac{1}{1-p_{\rm el}}.
\end{equation}
At the energy of SHiP, $p_{\rm el}\simeq0.25$ so that we estimate a mild increment of $\sim 30\%{}$.
In the following, we keep the conservative estimate based only upon the bremsstrahlung of the primary protons.  

\subsection{QCD prompt production}
\begin{figure}[t]
    \centering
    \includegraphics[width=0.25\textwidth]{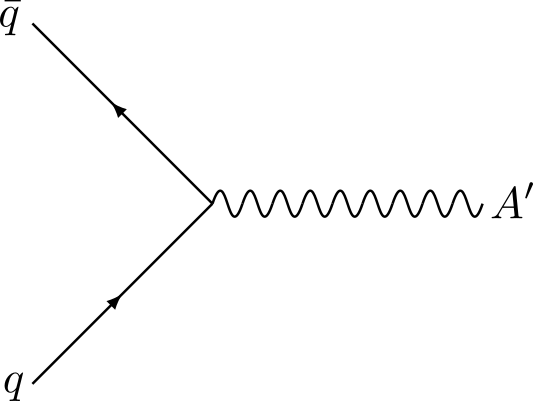}
    \hfill
    \includegraphics[width=0.3\textwidth]{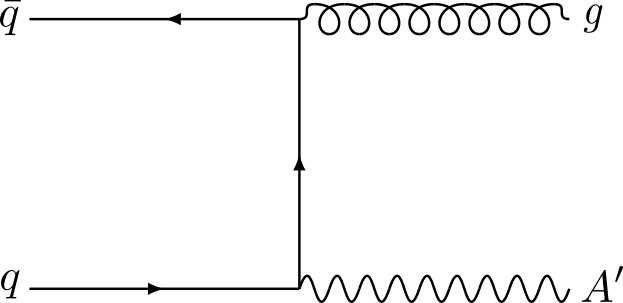}
    \hspace{0.15cm} \hspace{0.15cm} 
    \includegraphics[width=0.3\textwidth]{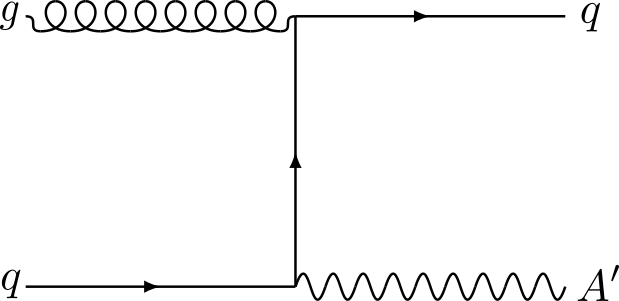}
    \caption{Main tree-level partonic QCD contributions: Drell-Yan-like production (left panel), associated production with the emission of extra QCD radiation (right panel).  }
    \label{fig:QCD-processes}
\end{figure}

QCD parton processes become relevant for $\mAp \gtrsim 1\,$GeV, at the edge where perturbation theory starts to become
reliable. Indeed, at scales $\lesssim 1\,$GeV the strong coupling $\alpha_s$ is $O(1)$ and the description of the hadrons in terms of constituent partons is spoiled by the confinement. In the attempt of estimating the relative importance
of this production mechanism, we have tried to push the perturbative computation down to $\mAp\sim 300\,$MeV.
The main tree-level diagrams correspond to the Drell-Yan-like production and the associated production with QCD radiation, Fig.~\ref{fig:QCD-processes}. 
The latter allows for smaller $\mAp$ values since the characteristic scale of the process, given by the transverse momentum of the emitted parton, can be kept
to be higher than the $\Lambda_\text{QCD}$ scale. A minimum cut on the $p_\mathrm{T}$ of the QCD radiation is physically required to tame infrared singularities. 
The cross section diverges logarithmically up to scales of order $O(\Lambda_\text{QCD})$, when perturbation theory eventually breaks down.
The transverse momentum cut-off is a severe requirement for an on-axis set-up as SHiP, due to its small angular acceptance. 
We find that even for relatively small values of the cut-off, $p_{\mathrm{T}}\sim 800\,$MeV, the production rate is not sufficient to produce a significant yield of LDM within the geometrical acceptance.
Therefore, we focus only on Drell-Yan-like production. We rely on \texttt{MadGraph5}(v.2.66)~\cite{Alwall:2011uj}, which is integrated in MadDump as it is based on the former package, for the generation of the events, and we use the \texttt{NNPDF2.3LO}~\cite{Ball:2012cx,Ball:2013hta} set as our choice of the proton parton distribution function (PDF). In the normalisation of the number of produced LDM particles, we effectively take into account nuclear effects in the following way
\[
N_{\mathrm{LDM}}=2\times\frac{\sigma_{pA\to\chi\bar{\chi}}}{\sigma_{pA\to\mathrm{all}}}\times N_{p.o.t.}=2\times\frac{A\,\sigma_{pp\to\chi\bar{\chi}}}{A^{0.71}\sigma_{pp\to\mathrm{all}}}\times N_{p.o.t.}=2\times A^{0.29}\times\frac{\sigma_{pp\to\chi\bar{\chi}}}{\sigma_{pp\to\mathrm{all}}}\times N_{p.o.t.}\,,
\]
where $A=A_{\mathrm{Molybdenum}}=96$; the nuclear rescaling as $A^{0.71}$ is taken from Ref.~\cite{ship} and $\sigma_{pp\to\mathrm{all}}=40\,\mathrm{mb}$~\cite{pdg2018}.\\
In this case, the characteristic scale of the process coincides with $\mAp$. As mentioned before, we cannot use scales $\lesssim 1\,$GeV, where both the strong coupling and PDF are ill-defined from the perturbative point of view. To push our projection into the sub-GeV range, we adopt the following prescription: we set both the re-normalisation scale $\mu_R$ (at which the strong coupling constant is evaluated) and the factorisation scale $\mu_F$ (at which the PDF is evaluated) to a fixed value chosen to be $\mu_R=\mu_F=1.5\,$GeV, the lowest scale variation point associated to open charm production.


\section{Background estimate}
\label{sec:Bkg}
\begin{figure}[htb]
     \centering
     \begin{subfigure}[b]{0.49\textwidth}
         \centering
         \includegraphics[width=\textwidth]{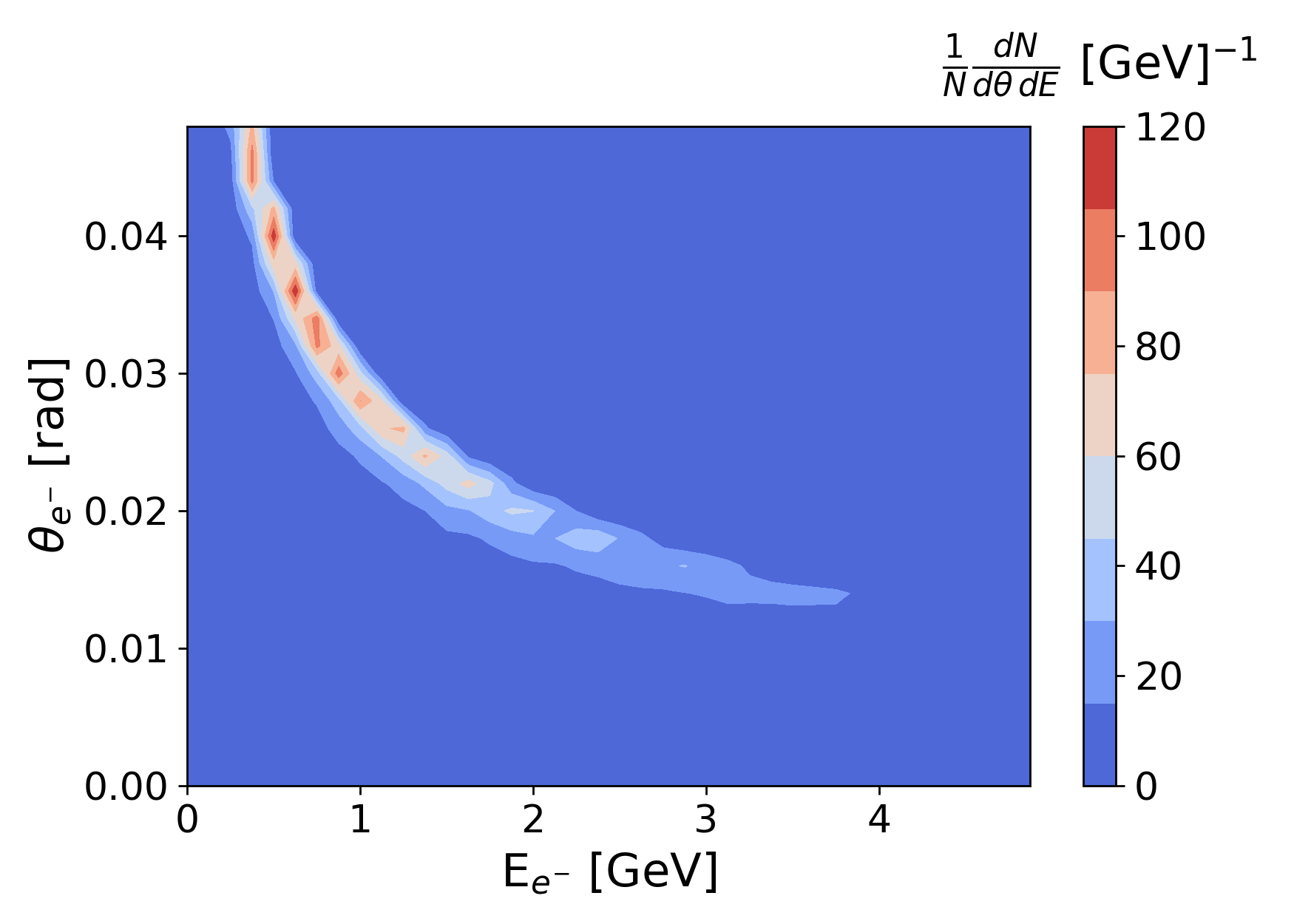}
         \caption{ $m_{A^\prime}$ = 50 MeV, production from $\pi^{0}$ decays.   }
         \label{subf:a}
     \end{subfigure}
     \hfill
     \begin{subfigure}[b]{0.49\textwidth}
         \centering
         \includegraphics[width=\textwidth]{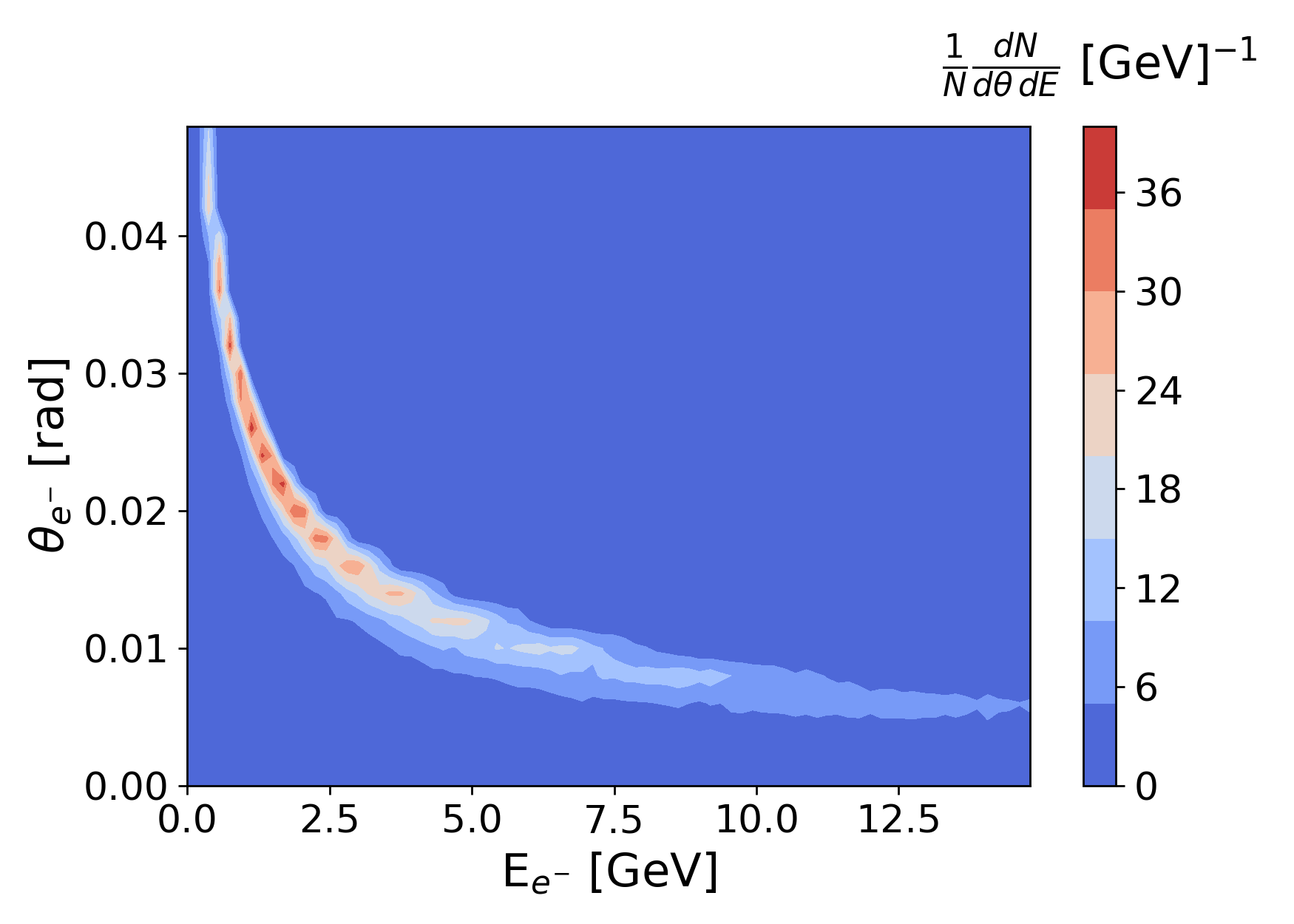}
         \caption{ $m_{A^\prime}$ = 250 MeV, production from $\eta$ decays. }
         \label{subf:b}
     \end{subfigure}
     \hfill
     \begin{subfigure}[b]{0.49\textwidth}
         \centering
         \includegraphics[width=\textwidth]{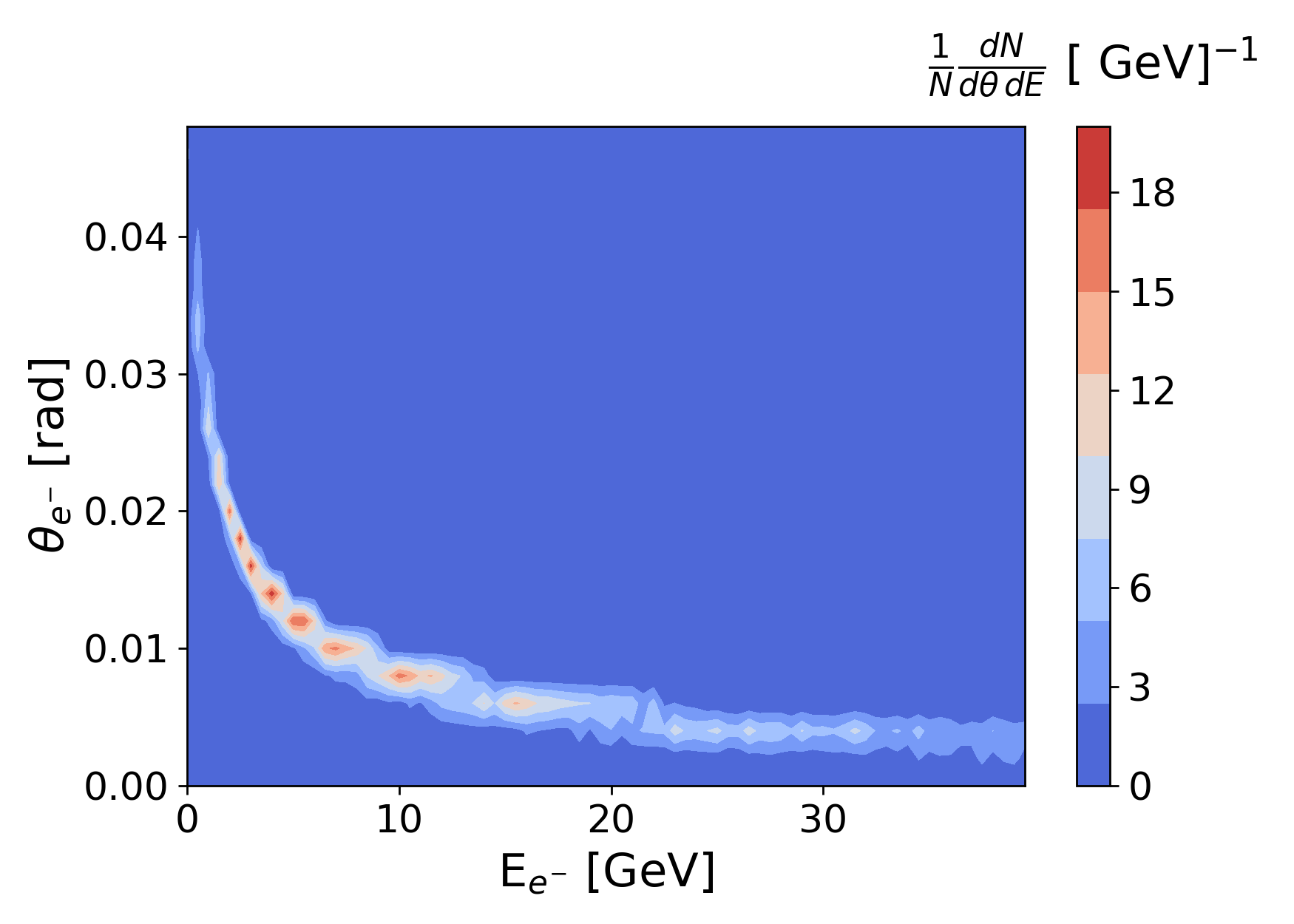}
         \caption{ $m_{A^\prime}$ = 500 MeV, production from proton bremsstrahlung. }
         \label{subf:c}
     \end{subfigure}
     \caption{2D-contour plot in the energy-polar angle plane of the recoil electron in LDM-$e^-$ scattering for three different mass DP candidates: (a) $50\,$MeV, (b) $250\,$MeV, (c) $500\,$MeV. The colour range is expressed in arbitrary units. A clear correlation is observed between the mass of the DP candidate and the electron energy-angle spectrum: the higher is the mass the smaller the recoil angle and the higher the associated energy. In the mass range we are interested in, most of the signal lies in the energy region below $10\,$GeV.  }
    \label{fig:signal2D}
\end{figure}

\begin{figure}[htb]
     \centering
     \begin{subfigure}[b]{0.49\textwidth}
         \centering
         \includegraphics[width=\textwidth]{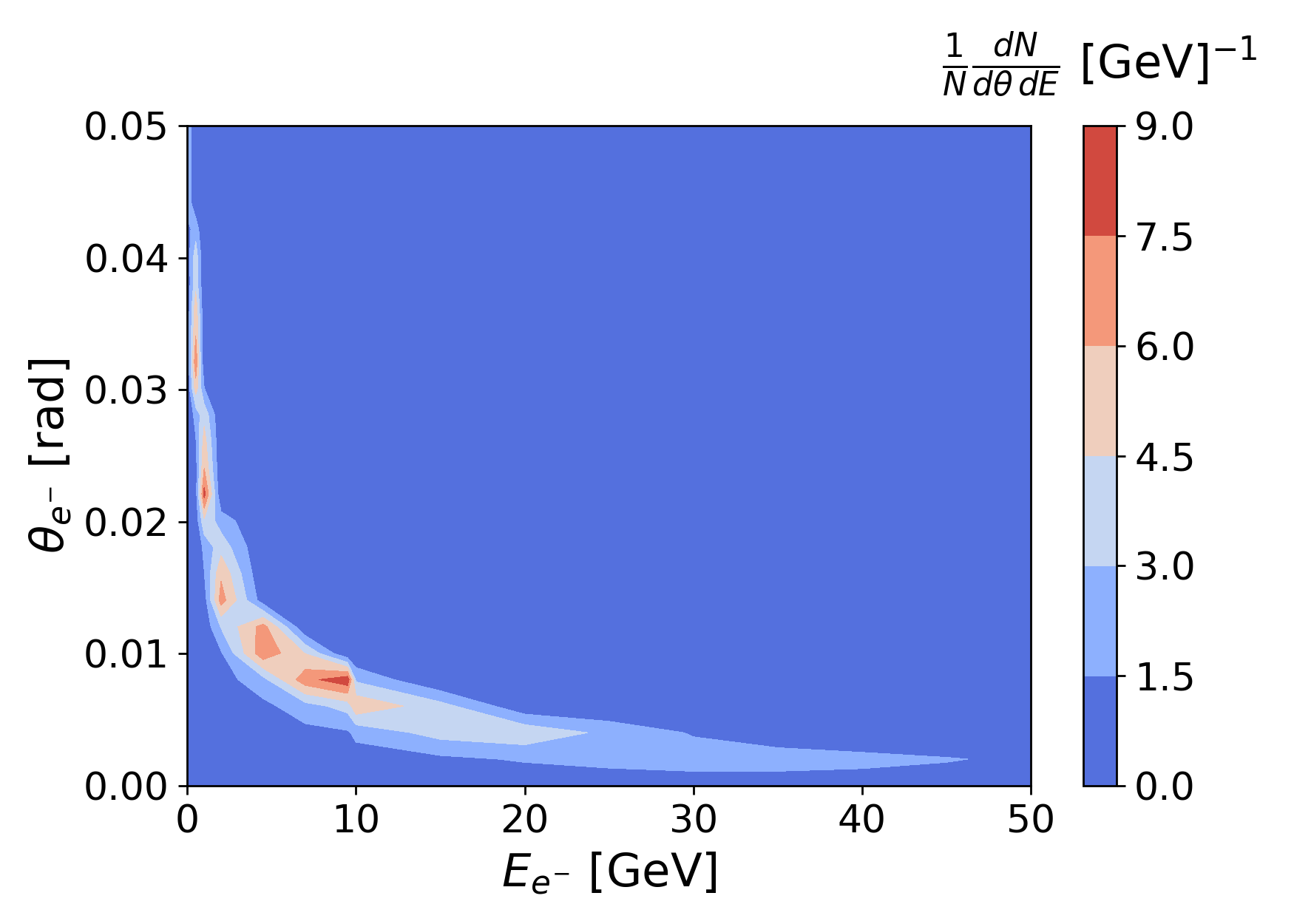}
         \caption{Sum of the EL $\nu_{\ell}\,(\bar{\nu}_{\ell})$ scattering contributions ($\ell=e,\,\mu$).}
         \label{subf:a}
     \end{subfigure}
     \hfill
     \begin{subfigure}[b]{0.49\textwidth}
         \centering
         \includegraphics[width=\textwidth]{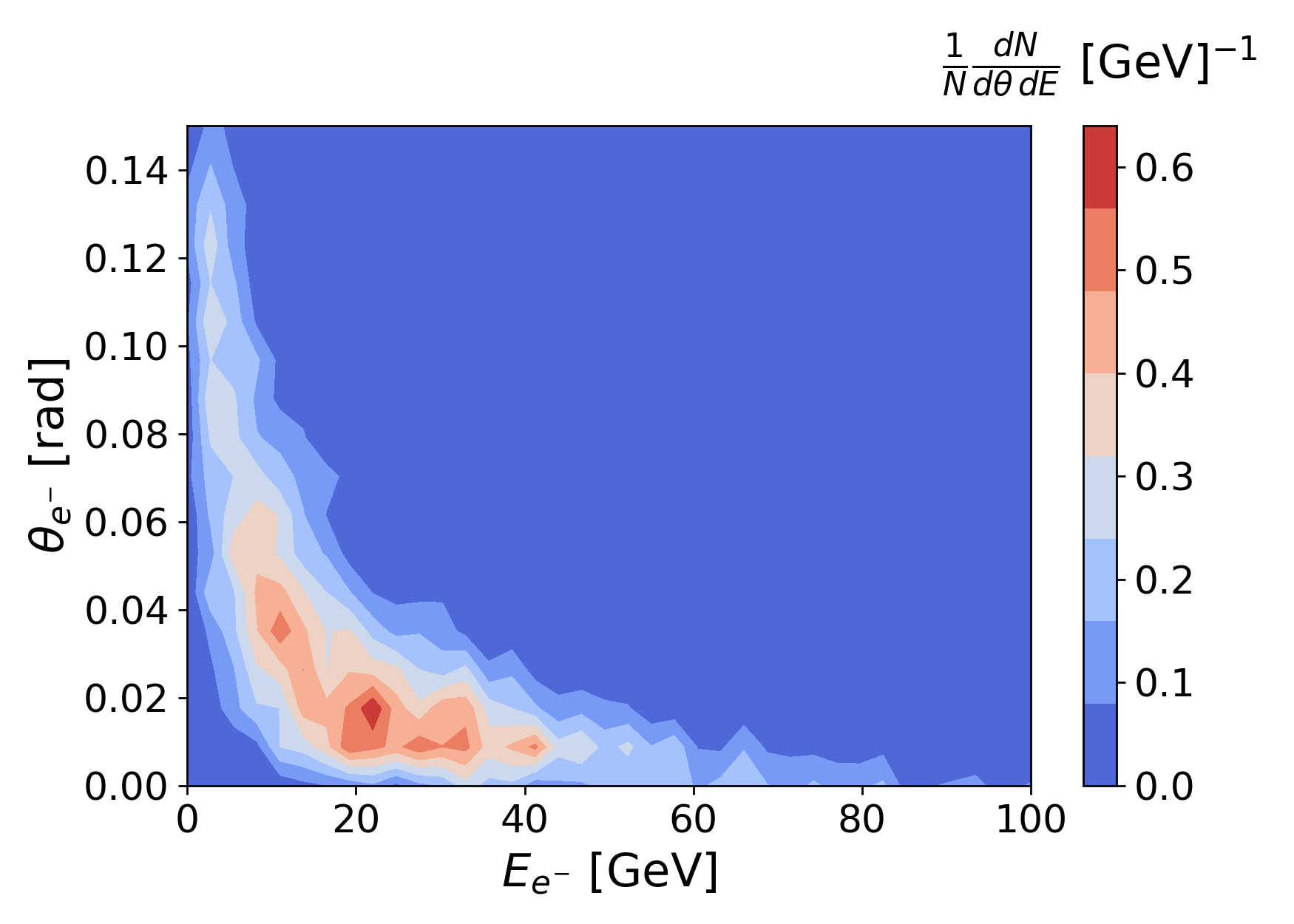}
         \caption{Sum of the QE $\nu_{e}(\bar{\nu}_{e})$ scattering contributions.}
         \label{subf:b}
     \end{subfigure}
     \caption{2D plot of the scattered electron energy $E_{e^{-}}$ Vs. angle $\theta_{e^{-}}$ for the relevant background sources from neutrino and anti-neutrino species: (a) EL scattering from $\nu_{\ell}\,(\bar{\nu}_{\ell})$ being $\ell=e,\,\mu$; (b) QE scattering from $\nu_e\,(\bar{\nu}_{e})$.}
    \label{fig:background2D}
\end{figure}

\begin{figure}[htb]
     \centering
     \begin{subfigure}[b]{0.49\textwidth}
         \centering
         \includegraphics[width=\textwidth]{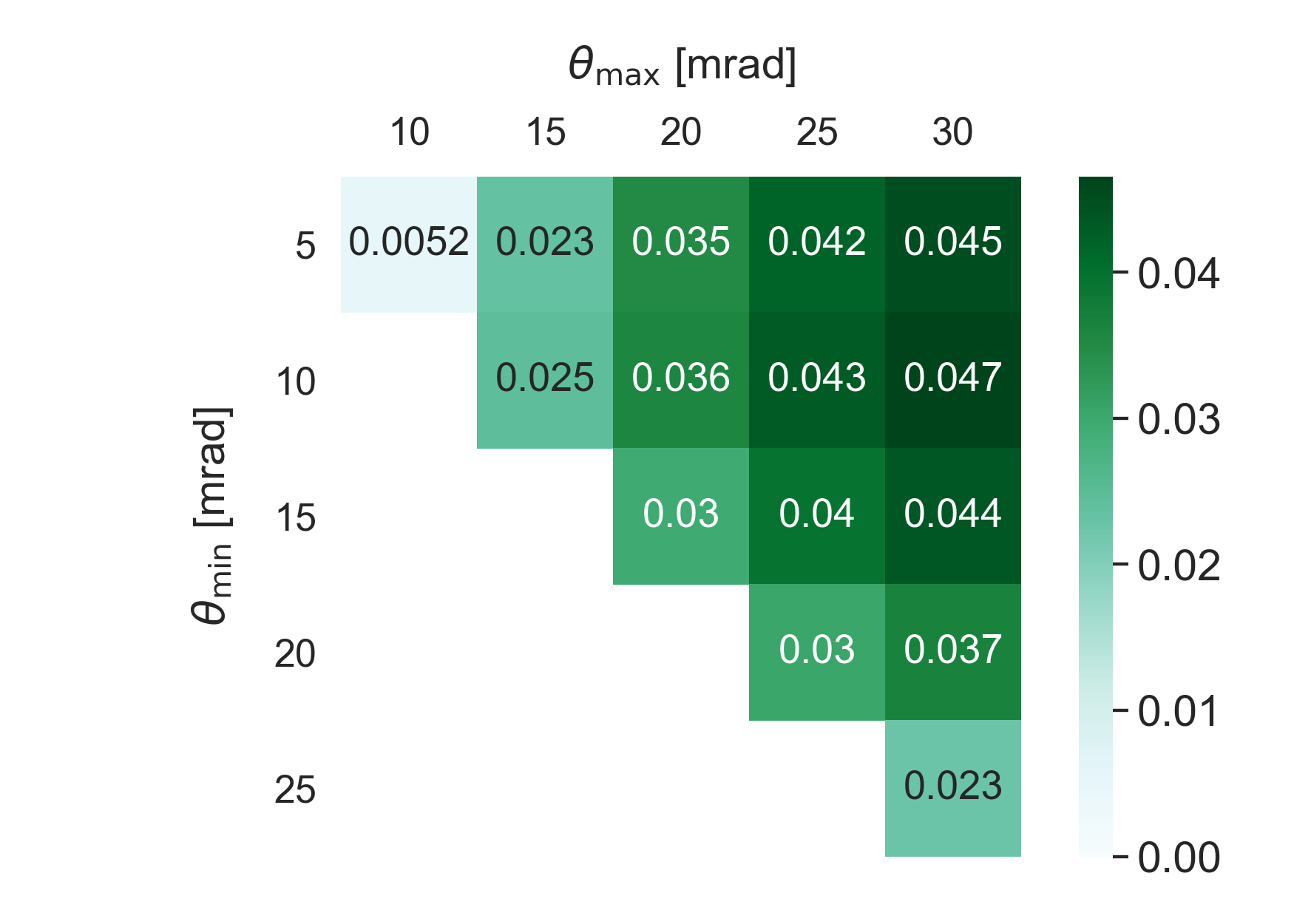}
         \caption{Grid values for $E_{e}\in[1,\,5]\,$GeV.}
         \label{subf:a}
     \end{subfigure}
     \hfill
     \begin{subfigure}[b]{0.49\textwidth}
         \centering
         \includegraphics[width=\textwidth]{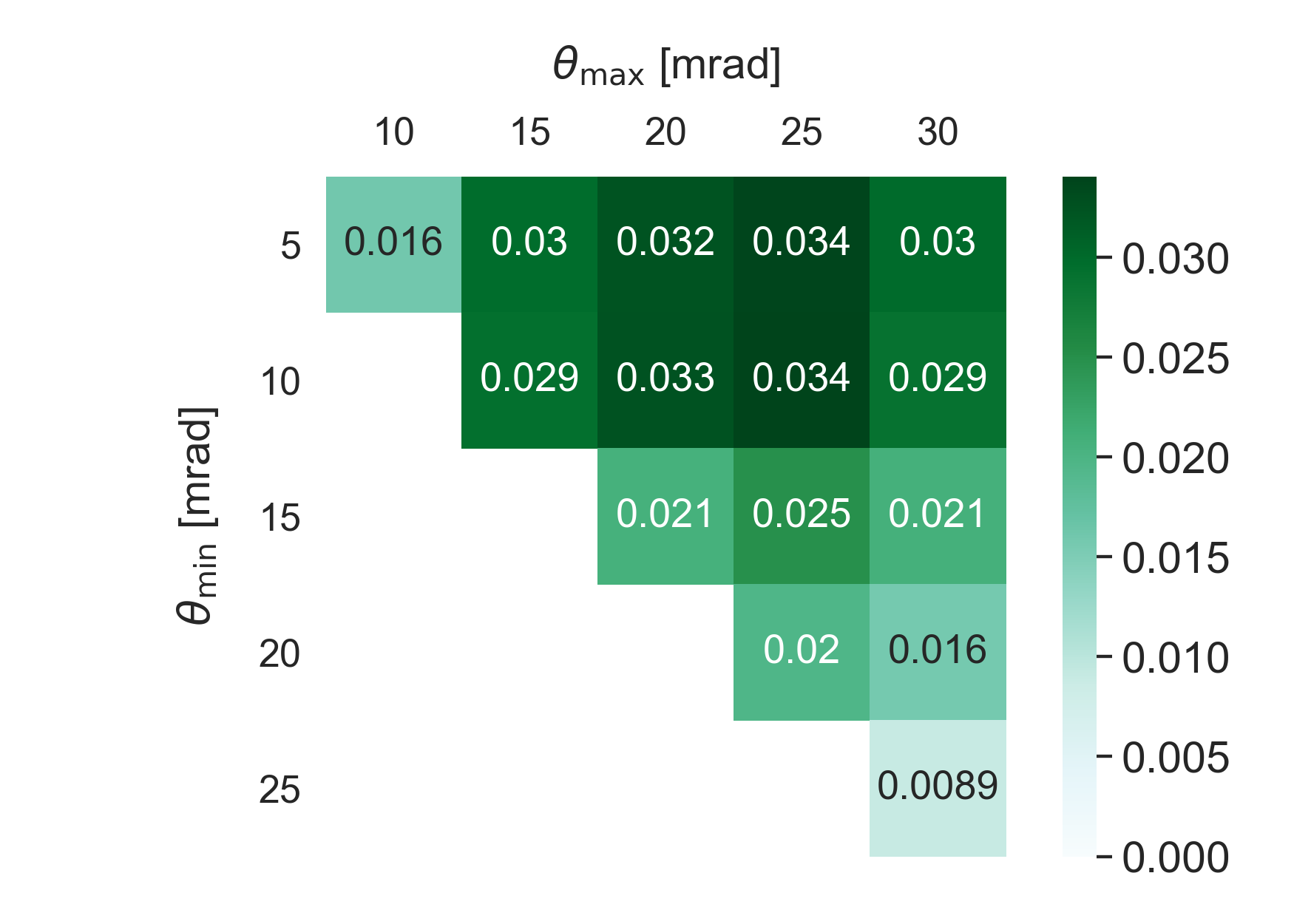}
         \caption{Grid values for $E_{e}\in[1,\,10]\,$GeV.}
         \label{subf:b}
     \end{subfigure}
     \caption{Grid-search optimisation of the significance $\Sigma$ as a function of the angular cut for a fixed energy window. The left axis represents the lower cut value for $\theta_e$ whereas the upper axis is the higher one. The plots in the two panels share the same normalisation. The best selection corresponds to the tighter energy window (a) and the angular range $[10,\,30]$ mrad.}
    \label{fig:gridopti}
\end{figure}

Neutrinos emerging from the beam-dump target and interacting in the SND are the relevant background source to the detection of LDM elastic scattering, whenever the topology at the primary vertex consists of a single outgoing charged track, an electron. The expected background yield for five years of data-taking has been estimated by means of the \texttt{GENIE}~\cite{ANDREOPOULOS201087} Monte Carlo software, supplied with the spectrum of neutrinos produced at the beam dump as simulated with \texttt{Pythia v6.4.28} within FairShip and including secondary production, for the generation of the following neutrino interactions in the whole kinematic phase space:
\begin{itemize}
    \item \textit{Elastic scattering} (EL) of $\nu_{e}(\bar{\nu}_{e})$,   $\nu_{\mu}(\bar{\nu}_{\mu})$ off the electrons of the SND, which is a source of irreducible background as it shares the same topology of LDM elastic interactions:
    \[
    \nu_{\ell}+e^{-}\to\nu_{\ell}+e^{-}\,.
    \]
    \item \textit{Resonant scattering} (RES) of $\nu_{e}(\bar{\nu}_{e})$ off nucleons A$(n,\, p)$:
    \[
    \nu_{e}(\bar{\nu}_{e})+A \to e^{-}(e^{+})+\Delta^{+/++}\,,
    \]
    \[
    \nu_{e}(\bar{\nu}_{e})+A \to e^{-}(e^{+})+(N^{*}\to\mathrm{inv})\,.
    \]
    \item \textit{Deep Inelastic scattering} (DIS) of $\nu_{e}(\bar{\nu}_{e})$ off nucleons A, representing background when only the electron track at the primary vertex is reconstructed because of unidentified hadrons:
    \[
    \nu_{e}(\bar{\nu}_{e})+A \to e^{-}(e^{+})+X\,.
    \]
    \item \textit{Quasi-elastic scattering} (QE) of $\nu_{e}(\bar{\nu}_{e})$, with the primary proton undetected because it is below the energy threshold:
    \[
    \nu_{e} + n \to e^{-} + p \, ,
    \]
    \[
    \bar{\nu}_{e} + p \to e^{+} + n \,.
    \]
\end{itemize}
Charged current interactions of $\nu_{\ell}\,(\bar{\nu}_{\ell})$ with $\ell=\mu,\,\tau$ do not represent a concern because they are easily discernible from LDM events by reconstructing the charged lepton produced in the final state. Electron decay modes of the $\tau$ lepton are a negligible background source, since an early decay of the parent track ($\sim 1\%$ occurrence) leading to an undetected $\tau$ would occur with less than a per-mill probability. In addition, we do not consider $\nu_{\tau}\,(\bar{\nu}_{\tau})$ elastic scattering processes as background, due to the suppression resulting from the combination of smaller flux $\phi_{\nu_{\tau}}$ ($\sim 1$ order of magnitude smaller than $\phi_{\nu_{e}}$ and $\sim 2$ orders of magnitude smaller than $\phi_{\nu_{\mu}}$) and cross section .\\
The whole $\nu$ spectrum is made to interact within the SND and the surrounding materials. Moreover, for this study we assume the detection efficiency to be unitary~\cite{Alexandrov:2015qkw}. 
\indent
The simulated sample of neutrinos undergoes a two-steps selection procedure, in order to be tagged as residual background.\\
\indent
First, only interactions occurring within geometrical acceptance and associated with a single charged final state track, an electron, are selected: $\nu$ vertices are further considered in the analysis only if located inside the SND volume, whereas all the out-coming charged tracks are inspected in order to assess their visibility in the nuclear emulsion medium. The visibility threshold depends crucially on the exploited tracking device technology; for this study we assume 170 MeV/c for the protons, 100 MeV/c for the other charged particles including the electrons. These are derived as benchmark values from the OPERA experiment, where charged-particle reconstruction is possible only if two consecutive straight track segments, before and after a lead plate, are found to be in agreement~\cite{ref:operavis}. A further handle considered here for signal against background discrimination is the presence of neutral particles, \textit{e.g.} photons or $\pi^{0}$s, nearby the interaction vertex, since it is not foreseen in any LDM elastic scattering event.\\
\indent
The second step of the event identification procedure consists of a kinematic selection in the energy $E_{e}$ and polar angle with respect to the incoming neutrino/LDM direction $\theta_{e}$ of the scattered electron. For the elastic case, these quantities are constrained by the kinematic relation $E_{e}\,\theta_{e}^{2}\leq2\,m_{e}\,$, valid in the regime $E_{in}\gg m_e,m_\chi$, where $E_{in}$ is the energy of the incident neutrino/LDM particle.
In order to choose the energy and angle ranges for the selection, an optimisation procedure is performed, aiming at maximising the following significance:
\begin{equation}\label{eq:significance}
    \Sigma = \frac{S}{\sqrt{\sigma_{\rm stat}^2+\sigma_{\rm sys}^2}} = \frac{S}{\sqrt{ B + \sum\limits_{\substack{i \in [\rm EL,\,QE,\,RES,\,DIS]\\ \ell \in [\nu_e,\nu_\mu,\bar{\nu}_e,\bar{\nu}_\mu]}} \left(\kappa_{i\ell} B_{i\ell}\right)^2}}\,,
\end{equation}
where $S$ denotes the signal yield, while $B_{i\ell}$ are the individual contributions to the background yield $B$ per interaction category and neutrino flavour, each of them weighted by a factor $\kappa_{i\ell}$ accounting for the systematic uncertainty. We have focused on the relevant systematics, arising from the uncertainty on the neutrino cross sections (assumed flavour-independent, $\kappa_i$) and on the incoming neutrino flux produced at the beam dump (interaction-independent, $\tilde{\kappa}_\ell$),  so that we have assumed $\kappa_{i\ell} = \sqrt{\kappa_{i}
^2+\tilde{\kappa}_\ell^2}$. As for the former, we assume the following: $5\%$ for DIS~\cite{Wu:2007ab}, $18\%$ for RES~\cite{PhysRevD.83.052007}, $8\%$ for QE~\cite{Lyubushkin:2008pe}, while we neglect the uncertainty on the EL cross section that is well known within the SM~\cite{RevModPhys.84.1307}.
As for the uncertainty on the incoming neutrino flux, this will be well constrained by an independent measurement of the abundant CC-DIS interactions occurring within the SHiP detector (expected $\sim\,10^6$ for $\nu_{e,\,\mu}$). Since the corresponding cross section is lepton-universal and known within $\sim 5\%$ accuracy down to $E_{\nu}$ of 2.5 GeV~\cite{Wu:2007ab}, we assume it to be the driving systematic uncertainty on the neutrino flux. While SHiP is capable of disentangling $\nu_\mu$ from $\bar{\nu}_\mu$ interactions by measuring the charge of the primary muon, thus providing a different estimate for $\nu_\mu$ and $\bar{\nu}_\mu$ fluxes, with regard to electron species it will measure a combination of the lepton and anti-lepton initiated events. As the relative abundance of $\nu_e$ and $\bar{\nu}_e$ produced at the beam dump can be assessed, the individual fluxes can be estimated accordingly. For neutrino energies below $2.5\,$GeV we double the uncertainty on the flux assuming them to be at $10\%$.\\
Since the signal yield $S$ depends on the mass hypothesis placed on the LDM candidate and thus on the DP, we adopt the most-general assumption of maximising the experimental sensitivity with respect to the broadest possible range of masses. Therefore, $S$ is given as the average of the signal yields for three DP mass hypotheses: $50\,$MeV, $250\,$MeV and $500\,$MeV.\\
The selection optimisation strategy is based on a grid-search method and proceeds as follows:
\begin{itemize}
    \item an energy window $[E_{\rm min},\,E_{\rm max}]$ is identified, according to the signal events distributions;
    \item in the given energy range, the significance $\Sigma$ values are determined  in uniform angular intervals of $5\,$mrad spread;
    \item the selection ranges, corresponding to the maximum $\Sigma$, are chosen for the analysis.
\end{itemize}
As shown in Fig.~\ref{fig:signal2D}, signal events are mostly concentrated at energies below $10\,$GeV. Two energy windows have thus been considered: $[1,\,5]\,$GeV and $[1\,,10]\,$GeV, where the lower cut is placed as a minimum requirement for the recoil electron to produce a detectable electromagnetic shower within the ECC brick. The motivation to consider an additional tighter energy range resides in the opportunity to further suppress the high energetic components of the neutrino background, as illustrated in Fig.~\ref{fig:background2D} which shows the relevant EL and QE contributions. DIS and RES processes are not shown since they exhibit signatures with higher multiplicities of charged tracks at the primary vertex.\\
The results of the optimisation are reported in Fig.~\ref{fig:gridopti}, showing indeed a preference for the tighter energy window $E_{e}\in[1,5]\,$GeV and an angular range $\theta_{e}\in[10,\,30]\,$mrad.\\
The corresponding background yield estimate is reported in Table~\ref{nuyield}.\\
\indent
Neutrino elastic scattering processes, involving either electronic and muonic species, represent the dominant background source and are to some degree irreducible, since they share the same topology as the signal.\\
With regard to quasi-elastic $\nu_{e}$ and $\bar{\nu}_{e}$ interactions, a small but non-negligible contribution is observed. The process $\nu_{e}\,n \to e^{-}\,p$ mimics the signal when the proton at the primary vertex is not identified, because of the 170 MeV/c threshold. Improvements in the proton identification efficiency with dedicated techniques, including Machine Learning clustering algorithms, will be the subject of future studies. When considering anti-neutrinos, events as $\bar{\nu}_{e}\,p \to e^{+}\,n$ are topologically irreducible since we assume for the present study the neutron to be undetectable within the SND. This effect compensates the larger (by a factor of $\sim 3$) neutrino flux, thus making the two contributions comparable.\\
\begin{table}[t]
\centering
\begin{tabular}{lccccc}
\hline
& $\nu_{e}$ & $\bar{\nu}_{e}$ & $\nu_{\mu}$ & $\bar{\nu}_{\mu}$ & all\\ 
\hline
Elastic scattering on $e^{-}$ & 68 & 41 & 60 & 38 & 207 \\
Quasi~-~elastic scattering  & 9 & 9 &  &  & 18 \\
Resonant scattering  & - & 5 &  &  & 5 \\
Deep inelastic scattering  & - & - &  &  & -\\
\hline
Total & 77 & 55 & 60 & 38 & 230 \\
\hline
\end{tabular}
\caption{Expected neutrino background yield to light dark matter elastic scattering search in the SHiP experiment, corresponding to $2\times 10^{20}$ delivered \textit{p.o.t.} The current estimate is the result of a combined geometrical, topological and kinematical selection, aimed at identifying only interactions occurring within the Scattering Neutrino Detector with one visible track in the final state being an electron. Tracks under a defined visibility threshold are discarded ($p<100$ MeV/c for charged, $p<170$ MeV/c for protons). A kinematic cut in $E_{e}\in [1,\,5]\,$GeV and $\theta_{e}\in [10,\,30]\,$mrad of the scattered electron is chosen as result of the signal significance optimisation procedure and determines the final number of background events. We refer to the Sec.~\ref{sec:Bkg} for further details on the analysis and the associated uncertainties.}
\label{nuyield}
\end{table}
In the case of resonant neutrino scattering, the outgoing electron is often accompanied by a further charged track, which helps discriminating between background and signal. Nevertheless, some topologically irreducible interactions are present as well:
\[
\bar{\nu}_{e}\,p\to e^{+}N^{*}\, ,\qquad N^{*}\to\Lambda^{0}\,K^{0}_{L/S}\,,
\]
where the $K^{0}_{L/S}$ is considered undetectable within the SND for this study. Future improvements lie in the employment of combined information of ECC and TT, coming from the linking of the emulsion tracks with those reconstructed in the electronic tracking system. Moreover, some final states with the pattern $e^{+}(n)\,\gamma$ contribute, when the emitted photon is too soft to be identified via the reconstruction of the electron-positron pairs from its conversion.\\
The contribution from neutrino deep inelastic scattering processes is, on the contrary, negligible, as a consequence of the high rejection power observed on these event categories, which exhibit a topology with a high multiplicity of charged tracks.\\
In the eventuality of an observed excess in the number of events, SHiP may collect data in a bunched beam mode, exploiting the time of flight measurement to separate massive particles like LDM from neutrinos.
\section{Sensitivity}
\label{sec:Sensitivity}
Once the significance of Eq.~\eqref{eq:significance} is maximised, the optimal energy and angle ranges are employed to determine the yields of signal and background, following a cut-and-count procedure, per each fixed value of the mediator mass $\mAp$. The $90\%\,$ confidence level (C.L.) exclusion limits on the $\epsilon$ coupling at fixed mass $\mAp$ are then retrieved by adopting a single-tail Poissonian statistics. Statistical and systematic uncertainties are combined as reported after Eq.~\eqref{eq:significance}.

\begin{figure}[t]
    \centering
    \includegraphics[width=0.95\textwidth]{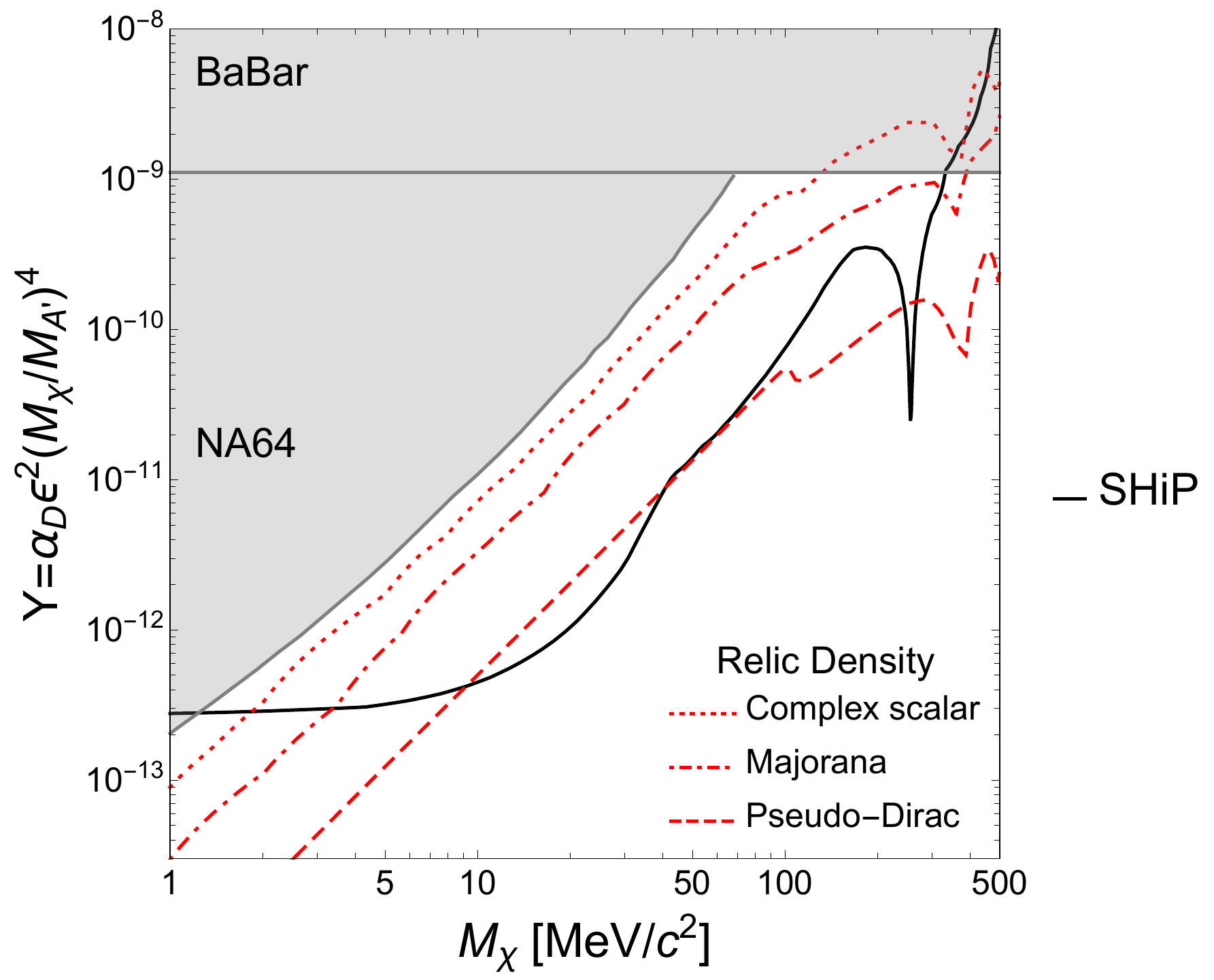}
    \caption{SHiP SND exclusion limit at $90\%\,CL$ relative to a $\Ap$ decaying into $\chi\bar{\chi}$ pairs for the benchmark point $\alpha_D= 0.1$ and $\mAp= 3\,m_\chi$. The current strongest experimental limits are also shown (BaBar~\cite{Lees:2017lec}, NA64~\cite{NA64:2019imj}), together with the three thermal relic lines corresponding to the scalar and the Majorana~\cite{cosmicvision}, and the Pseudo-Dirac DM~\cite{Duerr:2019dmv} hypothesis.
    }
    \label{fig:result}
\end{figure}

\begin{figure}[t]
    \centering
    \includegraphics[width=0.75\textwidth]{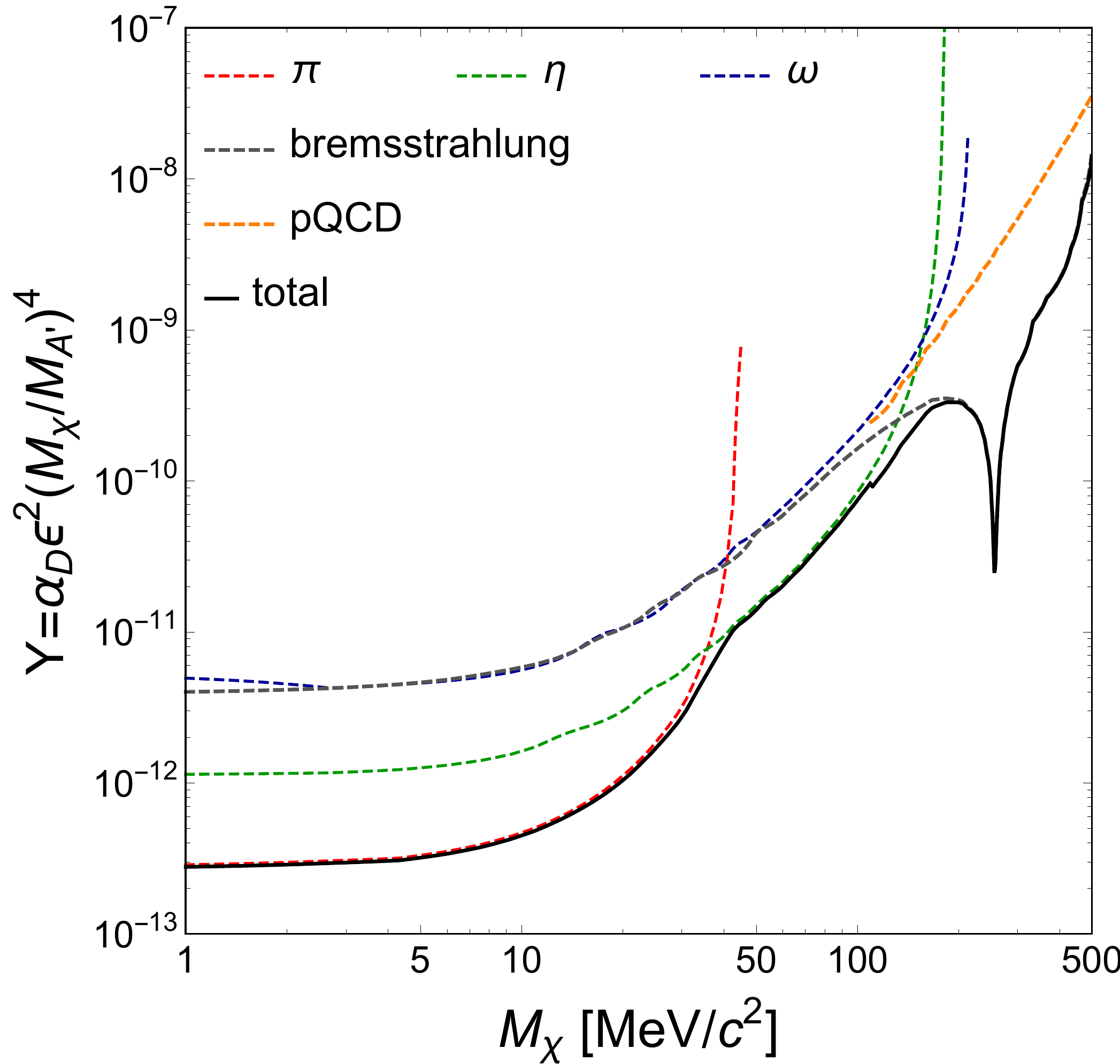}
    \caption{SHiP SND exclusion limit at $90\%\,CL$ relative to a $\Ap$ decaying into $\chi\bar{\chi}$ pairs for the benchmark point $\alpha_D= 0.1$ and $\mAp= 3 m_\chi$. The contributions from the different production mechanisms are reported separately.}
    \label{fig:result-contrib}
\end{figure}

In Fig.~\ref{fig:result}, we report our projection for the SHiP SND exclusion limit at $90\%$ C.L. in the $m_\chi-Y$ plane of the dark-photon model.
As stated above, we consider the benchmark scenario $\alpha_D=0.1$ and $m_\chi = {\mAp}/{3}$.
In Fig.~\ref{fig:result-contrib}, we separate the contributions given by the different production mechanisms.
In the low mass range $m_{\chi} \lesssim 150\,$MeV, the main contribution comes from the decay of the lightest mesons. $\pi$ decays dominate the $\Ap$ yield up to masses
close to the $m_\pi \to \gamma \Ap$ kinematic threshold. When approaching this threshold, the decay rate rapidly closes due to the steep suppression given by the phase space factor and with further increasing $m_\chi$ mass the $\eta \to \gamma \Ap $ starts to dominate.
The contribution of the $\omega$ is subdominant in the whole available mass range, which justifies \textit{a posteriori} the fact that we do not include in our
analysis $\Ap$ production from decays of heavier meson like the $\eta^\prime$. 

We find that the contribution due to pQCD is very small in the mass region explored. By varying the factorisation scale in the range $800\,{\rm MeV} < \mu_F < 3\,$GeV, we estimate the uncertainty associated with missing higher orders to be about $15{}\%$ on the signal yield within acceptance. We believe that this is an underestimation of the uncertainty as at next-to-leading order the process starts to receive radiative corrections proportional to the strong coupling constant at a scale close to $\Lambda_{\rm QCD}$, and new production channels open. While we do not expect that this will lead to a sizeable impact on the sensitivity, neglecting it leads anyway to a conservative estimate of the signal; hence, we have not considered the contribution of pQCD in our final result.

In the mass range $1\,$MeV $ < m_\chi < 300\,$MeV, the SHiP upper limit 
fairly improves the current strongest experimental limits (BaBar~\cite{Lees:2017lec}, Na64~\cite{NA64:2019imj}), 
even by more than an order of magnitude in the central region ($5\,$MeV $ < m_\chi < 100\,$MeV).
In this range and for the benchmark point under investigation, 
SHiP will cover the still unexplored parameter space corresponding to the solution of the relic density
given by a scalar LDM. In the range $3\,$MeV $< m_\chi < 300\,$MeV, SHiP will reach the thermal target for a Majorana candidate. Furthermore, it will exceed the thermal target for a Pseudo-Dirac candidate for masses around $10\,$MeV $< m_\chi < 40\,$MeV.  \\ 
\noindent We notice that for $m_\chi \lesssim 5\,$MeV the SHiP line saturates. 
In this region, the dark matter mass starts to become negligible 
and the selection requirements affect similarly the signal and the background.  
The rise in the signal production rate due to a lower mass is then balanced by a smaller fraction of events passing the kinematics selection, leading to the observed flatten sensitivity in the small mass range. The distinctive peak at $m_\chi\simeq 257\,$ GeV corresponds to
the $\rho-\omega$ resonant region, which is effectively taken into account by the 
time-like proton form factors used in the modelling of the proton bremsstrahlung mechanism.

\begin{figure}[t]
    \centering
    \includegraphics[width=0.95\textwidth]{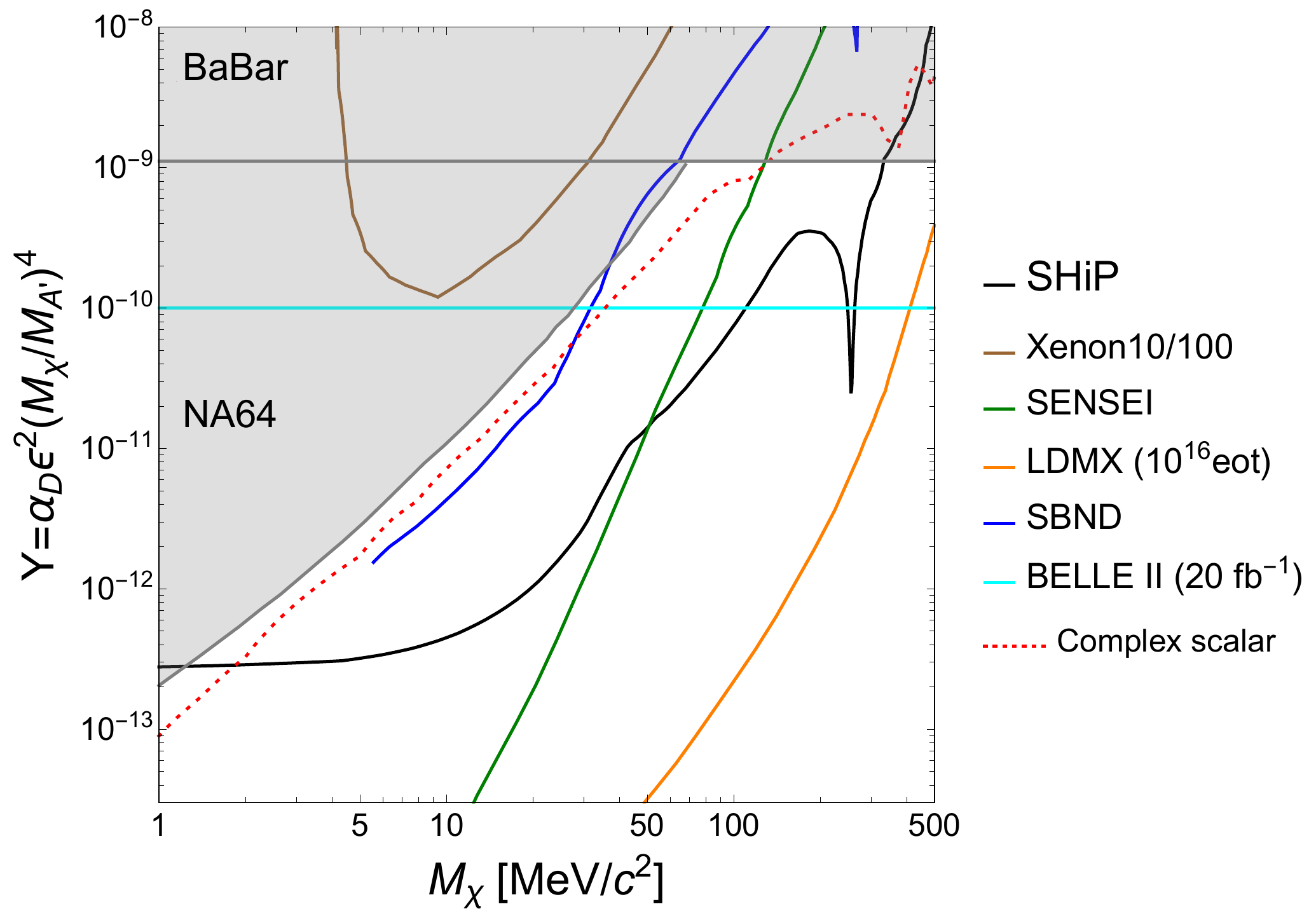}
    \hfill
    \includegraphics[width=0.95\textwidth]{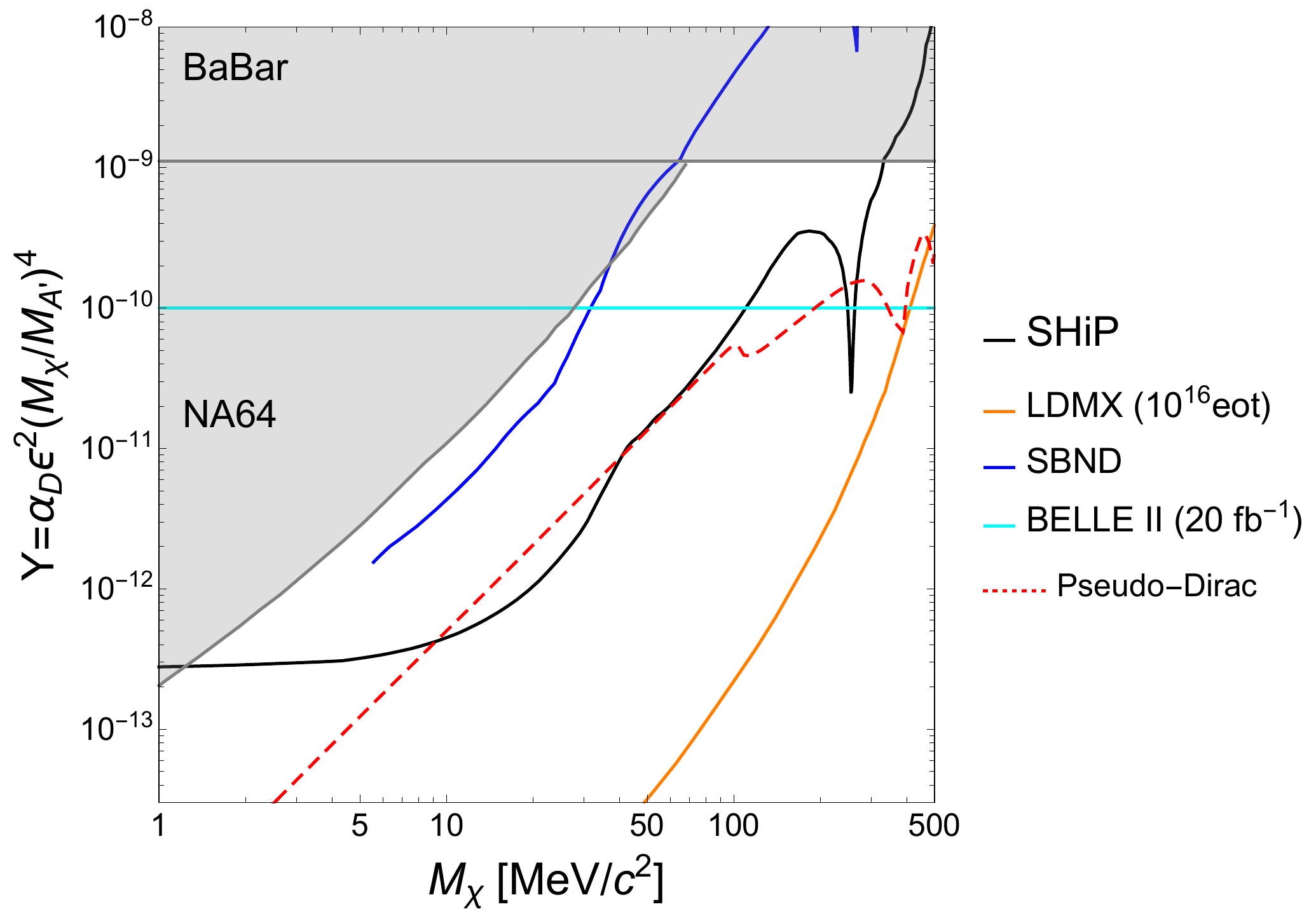}
    \caption{Comparison of existing and projected limits among the SHiP and other experiments as taken from Ref.~\cite{Alemany:2019vsk} for scalar (top panel) and Pseudo-Dirac dark matter (bottom panel).}
    \label{fig:result-cmp}
\end{figure}

In Fig.~\ref{fig:result-cmp}, the comparison between the SHiP sensitivity reach and that of other concurrent experiments clearly shows strengths and the complementarity offered by the proposed experimental scenario. Indeed the SHiP experiment will place constraints in unexplored regions of parameters space by exploiting a high intensity proton beam dump at $400\,$GeV and a micrometrical resolution tracking capability with the ECC. Thus, it offers a diverse approach to this NP search with respect to other experimental scenarios including direct searches and electron beam-line technologies.

\section{Conclusions}
\label{sec:Conclusions}
Light dark matter particles $\chi$ with masses in the sub-GeV region represent an appealing scenario for the explanation of the 
observed thermal relic density in the Universe. In this work, we have studied the potential offered by the SHiP SND to reveal LDM which couple to SM particles via a new gauge force mediated by a vector boson, $\Ap$.
We have assumed the simplest DP model, with coupling $g_D$ to $\chi$ and $\Ap$ kinetically mixed with the SM photon with mixing parameter $\epsilon$. We have 
focused on the relevant scenario for the SHiP SND: $ \mAp > 2 m_{\chi}$ and $ g_D \gg \epsilon e$.
Our main result is that for DM masses in $[1,300]\,$MeV the SHiP experiment will reach
an unexplored region of the parameter space. For the benchmark point considered, the sensitivity of the SHiP SND is even below the thermal relic line corresponding to a Majorana DM candidate in the mass window $[3,300]\,$MeV and it will reach the target for a Pseudo-Dirac candidate within $[15,30]\,$MeV.
Our analysis is based on a robust simulation framework for both the signal and the background which includes the relevant physical processes propagated within the detector. 
In particular, interactions of secondary particles in the beam-dump target have been taken into account in the neutrino background modelling, assuming unitary detection efficiency. As for the signal, we have consistently adopted $100\%$ detection efficiency within the selection requirements and we have studied the impact of the cascade effect on meson multiplicities and the resulting dark matter production yield. We have observed that the impact of the cascade is quite modest and
affects mainly the low masses. 

In this work, we have focused on the elastic $\chi\,e\to\chi\,e$ signature detectable within the SHiP Scattering and Neutrino Detector. Other signatures, as the elastic scattering with nuclei, may lead to an improvement of the sensitivity. We leave their study to forthcoming works. In our case, then the main background sources arise from 
elastic ${\nu_{\ell}}/{\bar{\nu}_{\ell}}$-electron and quasi-elastic ${\nu_{e}}/{\bar{\nu}_{e}}$ scattering. We have considered the region $E_{e}\in\,[1,\,5]\,$GeV and $\theta_{e}\in\,[10,\,30]\,$mrad, where $e$ is the recoil electron. We have found that about 230 neutrino events survive the selection requirements, for $2\times 10^{20}$ \textit{p.o.t.} corresponding to 5 years of data-taking.

We conclude by mentioning that, should an excess of events be observed,
a time of flight measurement of particles scattering within the SND might represent a smoking gun to discriminate LDM from neutrino events, thus leading to an inarguable discovery.

\section*{Acknowledgments}
The support from the Swiss National Science Foundation (SNF) under Contract No. BSSGI0\_155990, 200020\_88464 and IZSAZ2\_173357 is acknowledged.\\ 
The SHiP Collaboration wishes to thank the Castaldo company (Naples, Italy) for their contribution to the development studies of the  decay vessel. The support from the National Research Foundation of Korea with grant numbers of 2018R1A2B2007757, 2018R1D1A3B07050649, 2018R1D1A1B07050701, 2017R1D1A1B03036042,\\
2017R1A6A3A01075752, 2016R1A2B4012302, and 2016R1A6A3A11930680 is acknowledged. The support from the FCT - Funda\c{c}\~{a}o para a Ciência e a Tecnologia of Portugal with grant number CERN/FIS-PAR/0030/2017 is acknowledged. The support from the Russian Foundation for Basic Research (RFBR), grant 17-02-00607, and the support from the TAEK of Turkey are acknowledged.

\clearpage
\bibliographystyle{JHEP.bst}    
\bibliography{biblio,NOVA}
\end{document}